\documentclass[prb,showpacs,twocolumn,superscriptaddress]{revtex4-2}

\usepackage{amssymb}
\usepackage{graphicx}
\usepackage{bm}
\usepackage{xcolor}
\usepackage{epsfig}
\usepackage{amsfonts}
\usepackage{amsmath}
\usepackage{ulem}

\newcommand{\beqa}{\begin{eqnarray}}
\newcommand{\beeq}{\begin{equation}}
\newcommand{\eeqa}{\end{eqnarray}}
\newcommand{\eeqe}{\end{equation}}

\begin{document}

\title{Quantum Interference Control of Coupled Atomic Spins}

\author{Sergio Santiago Gomez}

\affiliation{Instituto de Modelado e
Innovaci\'on Tecnol\'ogica (CONICET-UNNE), Avenida Libertad 5400, W3404AAS
Corrientes, Argentina.}

\affiliation{Facultad de Ciencias Exactas, Naturales y Agrimensura, Universidad Nacional del Nordeste, Avenida Libertad 5400, W3404AAS
Corrientes, Argentina.}

\author{Gustavo Mauricio Meneses Ancalle}
\affiliation{Centro Atómico Bariloche and Instituto Balseiro (Universidad Nacional de Cuyo), 8400 San Carlos de Bariloche, Río Negro, Argentina}


\author{Daniel Dom\'inguez}

\affiliation{Centro Atómico Bariloche and Instituto Balseiro (Universidad Nacional de Cuyo), 8400 San Carlos de Bariloche, Río Negro, Argentina}

\author{María José Sánchez}

\affiliation{Centro Atómico Bariloche and Instituto Balseiro (Universidad Nacional de Cuyo), 8400 San Carlos de Bariloche, Río Negro, Argentina}

\affiliation{Instituto de Nanociencia y Nanotecnología, CONICET-CNEA, 8400 San Carlos de Bariloche, Rio Negro, Argentina}

\author{Alejandro Ferr\'on}
\affiliation{Instituto de Modelado e
Innovaci\'on Tecnol\'ogica (CONICET-UNNE), Avenida Libertad 5400, W3404AAS
Corrientes, Argentina.}

\affiliation{Facultad de Ciencias Exactas, Naturales y Agrimensura, Universidad
Nacional del Nordeste, Avenida Libertad 5400, W3404AAS
Corrientes, Argentina.}

\begin{abstract}

We propose a protocol for the non-resonant electric control of quantum states in atomic-scale spin systems using Landau-Zener-Stückelberg-Majorana (LZSM) interferometry. We focus on a Ti-Ti-Fe trimer on a thin MgO layer, a platform that has recently emerged as a promising candidate for qubit realization using electron spin resonance combined with scanning tunneling microscopy (ESR-STM). We first analyze a Ti dimer in the absence of the Fe atom and show, through a detailed spectroscopic study, that avoided level crossings can be engineered by combining the in-plane component of the external magnetic field, the tip magnetic field, and the applied bias voltage, which allows the system to be tuned electrically through these crossings. Based on this, we develop LZSM protocols for coherent state control in the dimer using non-resonant LZSM-STM. We then consider the full trimer, where the Fe atom generates a local magnetic field at the remote Ti atom, enabling non-local spin control. We show that this remote and non-resonant manipulation reduces unwanted effects associated with tunneling electrons and enables quantum operations on short timescales. Our results show that quantum interference can be used as a resource for electrically driven, non-resonant, and non-local spin control in atomic-scale systems. This approach enables fast quantum operations while mitigating decoherence, offering a promising route toward scalable atom-based qubit architectures.

\end{abstract}
\date{\today}

\maketitle

\section{Introduction}


Coherent control of spin systems is commonly achieved through resonant driving, which, in the experiments considered here, is implemented using ESR-STM \cite{Baumann2015,taherpour2026review-esr, willke2021a,wang2023qb,willke2018,willke2018b,willke2019,willke2019b,yang2017,seifert2019,seifert2021,seifert2020}. This technique enables coherent manipulation and spectroscopy of individual spins on surfaces by applying radio-frequency signals through the STM tip while measuring the tunneling current~\cite{Baumann2015,yang2017,yang2019,willke2018,willke2021a,chen2023harnessing}, providing atomic-scale access to their quantum properties.  However, resonant drivings are usually implemented with small signals whose amplitude determines the rotation frequency. Consequently, faster quantum operations require increasingly larger drive amplitudes.  As an alternative, Landau-Zener-Stückelberg-Majorana (LZSM) \cite{nori1,nori2,campbell_2020,caceres_2023,ryzhov2024,ferreyra2026optimizing} control offers a distinct non-resonant approach to achieving fast quantum operations. Instead of driving the system at the target transition frequency with a rather small amplitude, one periodically modulates the control parameter with a non-resonant periodic driving of a moderate (or even strong) amplitude, 
thereby sweeping the system through an avoided crossing. The resulting dynamics is governed by the coherent interference between probability amplitudes generated during successive passages through the avoided crossing \cite{nori1,nori2}. Both protocols enable the control of on-surface spin states and, more importantly, can be implemented using exactly the same experimental setup \cite{rodriguez2024non,wang2025electrically}.


The fundamental requirement for implementing the LZSM protocols considered here is an avoided crossing between two spin states in the energy spectrum, together with a means of sweeping the system through it. Titanium dimers on MgO/Ag surfaces offer a suitable platform for LZSM control. Their interatomic coupling can be tuned by adjusting the Ti–Ti distance during assembly \cite{yang2017,bae2018}. Kot and co-workers \cite{kot2023electric} studied their spin excitation spectrum under perpendicular magnetic fields and static electric fields from the STM tip. Depending on the interatomic separation, they find crossings between low-lying states as the external fields varied. However, they did not find an avoided crossing between the ground and first excited state in that experimental configuration, indicating that the coupling responsible for hybridization remains negligible under these conditions. 


In a different experimental approach, Wang and co-workers \cite{wang2025electrically} demonstrated that it is possible to access avoided crossings in Ti dimers using radio-frequency driving. They find an avoided level crossing by transforming to the rotating frame at a frequency $\omega_{rf} $. In the rotating frame, the oscillatory tip field becomes a static transverse magnetic field, which hybridizes the Ground State and the First excited state, opening an anticrossing at zero energy detuning. 


Recently, the spin of a remote atom has been successfully controlled \cite{wang2023qb,soohion2023b}. Here, “remote” refers to an atom that is not located directly below the STM tip and is therefore not exposed to the tunneling-electron current, one of the main sources of decoherence in ESR-STM. This scheme requires an Fe atom near a remote Ti spin so that the modulation of their exchange interaction can induce a spin flip on it. This approach has recently been used to construct and coherently control atomic qubits and even to implement quantum gates \cite{wang2023qb}. Remote control has also been proposed for topological chains, which could be implemented and controlled using the same experimental techniques discussed here \cite{YelkoSP2025}. Here, we show that remote control can be combined with LZSM protocols to manipulate a remote spin through quantum interference. We investigate the engineering of avoided crossings in a Ti dimer coupled to an Fe atom and use the electric field to achieve remote control via LZSM-STM.

We begin by examining a Ti dimer to show how avoided crossings can open and how their gaps and electrical tunability depend on the STM tip position and the external magnetic field. This minimal system also reveals a key limitation: although LZSM interference enables efficient manipulation of the spin beneath the tip, the weak electrical driving of the relevant transition severely constrains control of the remote spin.

Introducing an Fe atom near the remote Ti atom provides an additional, electrically tunable exchange interaction that overcomes this limitation. We show that this architecture supports avoided crossings suitable for remote spin control, with gaps that can be tuned through the external magnetic field. Voltage-induced modulation of the Ti–Fe exchange then drives repeated passages through these avoided crossings, enabling LZSM interference to selectively reverse the remote Ti spin while leaving the spin beneath the tip unchanged. 

\begin{figure}[hbt]
\includegraphics[width=1.0\linewidth]{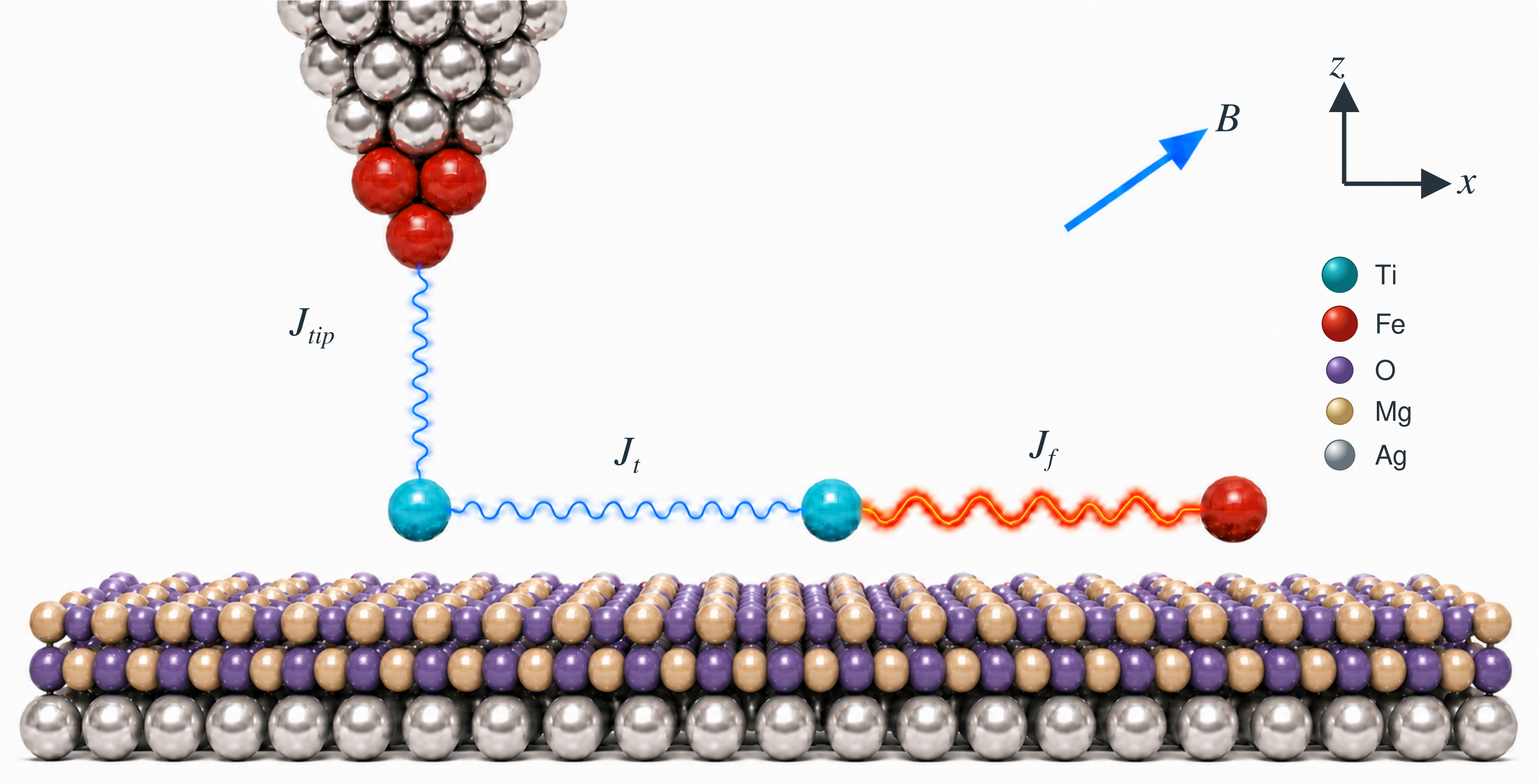}\\
\caption{\label{fig1} Sketch of a Ti-Ti-Fe complex in MgO/Ag and an SP-STM tip.}
\end{figure}

The paper is organized as follows. In Section~\ref{smod}, we introduce the model describing the atomic array deposited on MgO and its interaction with the STM tip (see Fig. \ref{fig1}). We also analyze the possible mechanisms for electrical quantum-state control and derive an effective model that describes the low-energy levels and the different mechanisms responsible for opening avoided crossings and controlling the system. In Section \ref{stidim}, we explore the possibility of engineering avoided crossings in  dimers and subsequently study the potential for controlling quantum states using LZSM quantum interference. Section \ref{sremote} demonstrates the possibility of implementing remote quantum control protocols based on LZSM interference by exploiting the modulation of the coupling between the Fe atom and the Ti atom located far from the STM tip. We analyze several possible implementations and show that remote control can be achieved on extremely short timescales. Finally, in Section \ref{sc}, we summarize the main results obtained and present our principal conclusions.

\section{Titanium dimer with a nearby iron atom}
\label{smod}
\subsection{Spin Hamiltonian for the trimer}

We model the Ti-Ti-Fe trimer on MgO using an effective spin Hamiltonian, following Refs.~\cite{lado2017,ferron2019,rodriguez2023optimizing,rodriguez2024non}:
\begin{equation}\label{htot}
H_{\rm S}= H_{\rm Ti}+ H_{\rm Fe} + H_{\rm int} + H_{\rm tip}.
\end{equation}

The Ti atoms (Spin $S_i=1/2$, $i=1,2$) are described by the Zeeman Hamiltonian,
\begin{equation}\label{hti}
H_{\rm Ti}= \mu_B \sum_{a=x,z} B_a \, g_{Ti}^{(a)} \left(S_1^{(a)}+S_2^{(a)}\right),
\end{equation}
where  $g_{Ti}^{(a)}$ are the components of the gyromagnetic tensor for the Ti atoms, and $B_a$ are the components of the external magnetic field.

The Fe atom ($S_{\rm Fe}=2$) includes both Zeeman and single-ion anisotropy terms \cite{rodriguez2023optimizing, rodriguez2024non},
\begin{equation}\label{hfe}
H_{\rm Fe}= -{\cal D}_2 (S_{\rm Fe}^{(z)})^2 + \mu_B \sum_{a=x,z} B_a \, g_{Fe}^{(a)} S_{\rm Fe}^{(a)}.
\end{equation}

\noindent Here, ${\cal D}_2$ is the axial zero-field splitting, which provides the dominant magnetic anisotropy and defines the quantization axis along $z$. This term produces a large energy separation between the lowest doublet, with $m_{\rm Fe}=\pm 2$, and the higher-energy levels. The coupling between the two states in this doublet is very weak  and can be neglected in the presence of an external magnetic field.

The exchange interaction is modeled as
\begin{equation} \label{hint}
H_{\rm int}= J_t\, {\bf S}_1 \cdot {\bf S}_2 + J_f\, {\bf S}_2 \cdot {\bf S}_{\rm Fe},
\end{equation}
where $J_t$ and $J_f$ denote Ti-Ti and Ti-Fe couplings, respectively. Dipolar interactions are neglected, since they are significantly smaller than the exchange in the regime of interest \cite{kot2023electric,phark2023electric}.

The effect of the STM tip is treated semiclassically as an effective local magnetic field acting on the first Ti atom \cite{lado2017,rodriguez2023optimizing,rodriguez2024non},
\begin{equation}\label{htip}
H_{\rm tip}= \mu_B \sum_{a=x,z} g_{Ti}^{(a)}\, b_a \, S_1^{(a)}.
\end{equation}
The tip-induced field $\mathbf{b}$ includes both dipolar and exchange contributions \cite{rodriguez2023optimizing}. Its components can be written as
\begin{align}
b_z &= 2\left(\frac{J_{\rm tip}(z)}{\mu_B g_{Ti}^{(z)}} - \frac{\mu_0 M_{\rm tip}}{4\pi |z|^3}\right)\sin(\theta+\eta),\label{b0z} \\
b_x &= 2\left(\frac{J_{\rm tip}(z)}{\mu_B g_{Ti}^{(x)}} + \frac{\mu_0 M_{\rm tip}}{8\pi |z|^3}\right)\cos(\theta+\eta),\label{b0x}
\end{align}
where $M_{\rm tip}$ is the magnetic moment of the tip, $z$ is the tip-Ti distance, and $\theta$ is the angle of the external magnetic field with the surface. Due to the tip magnetic anisotropy, the tip magnetization can be misaligned from the external field by an angle $\eta$ \cite{rodriguez2023optimizing,ferron2019}. The exchange coupling is given by $J_{\rm tip}(z)=J_{0}\,e^{-z/l_{0}}=J_0^\ast\, e^{-(z-z_c)/l_{0}}$, where $J_0$, $l_0$, and $z_c$ depend on the atom located directly beneath the STM tip, and $z_c$ defines a reference distance known as the point contact \cite{willke2019,rodriguez2023optimizing}. The model parameters used in this work are summarized in Table~\ref{table1}, together with the references supporting their choice.

\begin{table}[t]
\centering
\caption{Model parameters used in this work.
The exchange factor $J_0$ is estimated using results reported for the resonance frequency in \cite{willke2019}. The effective exchange prefactor $J_0^\ast$ is estimated for a Ti atom using $z_c=0.3$--$0.4$ nm \cite{willke2019,rodriguez2023optimizing}.}
\label{table1}
\begin{tabular}{@{}lll@{}}
\toprule
Parameter & Value & References \\
\hline
${\cal D}_2$
& $5$ meV & \cite{rodriguez2023optimizing,rodriguez2024non} \\
$J_0$, $J_0^\ast$ & $4$ eV, $0.25$ meV 
& \cite{willke2019} \\
$z_c$ & $0.3$--$0.4$ nm & \cite{willke2019} \\
$l_0$ & $0.043$ nm & \cite{rodriguez2023optimizing,ferron2019,willke2019} \\
$M_{\mathrm{tip}}$ & $\sim 30\,\mu_B$ & \cite{rodriguez2023optimizing,willke2019} \\
\hline
\end{tabular}
\end{table}


Also, in our calculations we set $g_{Fe}^{(z)}=2.8$ and $g_{Ti}^{(z)}=g_{Ti}^{(x)}=2.0$.  If the anisotropy of the $g$-factor is taken into account ($g_{Ti}^{(z)}\simeq1.9$, and $g_{Ti}^{(x)}\simeq2.0$), the qualitative results remain the same, although small shifts of the energy levels are expected. These shifts arise because the anisotropic $g$-tensor modifies the effective magnetic field experienced by the Ti spins, in particular its orientation with respect to the principal axes of the system. 


\subsection{Quantum Control using the electric field}

Applying an electric field with the STM tip induces a strain $\delta z$ in the bond between the adatom and the underlying surface \cite{lado2017}. In the present case, Ti atoms can occupy either the oxygen site of the MgO bilayer or a bridge position between two oxygen atoms \cite{willke2018b,yang2017,yang2019}. The effect of an electric field on the atom is to change its position with respect to the surface \cite{lado2017,ferron2019,rodriguez2023optimizing}. The electric field at the tunnel junction can be approximated as $E = V(t)/z_{tip}^{(0)}$, where $z_{tip}^{(0)}$ is the equilibrium tip-adatom distance. The applied bias voltage, in usual resonant and non-resonant experiments, contains both dc and rf components and is written as
\begin{equation}
    V(t)=V_{\mathrm{dc}}+V_{\mathrm{rf}}\sin(\omega t),
    \label{voltage}
\end{equation}
where $V_{\mathrm{dc}}$ determines the static operating point,
$V_{\mathrm{rf}}$ is the driving amplitude, and $\omega$ is the driving
frequency. This field exerts a force on the adatom, $F \simeq q V(t)/z_{tip}^{(0)}$, which is balanced by a restoring elastic force $F = -k_{\text{at}} \delta z$. Therefore, the displacement reads \cite{lado2017}:
\begin{equation}
    \delta z(t) \simeq \frac{q V(t)}{k_{\text{at}} z_{tip}^{(0)}},
    \label{piezo}
\end{equation}
where $q=1$ for Ti atoms. Estimates based on DFT yield displacements of order $\sim 1$ pm for $V \sim 100$ mV and $z_{tip}^{(0)} \sim 0.4$ nm \cite{ferron2019,yang2019,kot2023electric}. However, experimental observations suggest larger effective displacements. In practice, using values of $k_{\text{at}}$ in the range $50$–$150$ eV/nm$^2$, we reproduce the experimentally observed energy shift, defined as the shift in the resonance frequency induced by the STM tip \cite{willke2019}. These values are approximately 2 to 6 times smaller than those obtained using DFT, indicating a possible underestimation of the atomic displacement or the presence of additional mechanisms other than piezoelectric displacement. 

The displacement $\delta z(t)$ makes the Hamiltonian parameters time-dependent. In particular, the magnetic field generated by the tip depends on the instantaneous distance $z = z_{tip}^{(0)} - \delta z(t)$, leading to a modulation of $b_x$ and $b_z$ as described in Eqs.~(\ref{b0z}) and~(\ref{b0x}).


The electric field can also modulate the exchange interaction between two atoms, as discussed in previous works \cite{soohion2023b,wang2023qb,phark2023double}. We model the tunnel-junction electric field as that generated by a parallel-plate capacitor formed by the metallic tip and the substrate \cite{wang2023qb}. Each atom experiences a local electric field whose effect depends on its atomic species and adsorption environment. The resulting relative displacement between the atoms modifies their exchange coupling \cite{soohion2023b}. For equivalent atoms, i.e., atoms of the same species in identical local environments, this relative displacement, and hence the modulation of their exchange interaction, is expected to be negligible. In our system, the Ti–Ti exchange interaction, therefore, remains essentially unchanged, whereas the coupling between the Fe atom and the second Ti atom can be significantly modulated. An alternative mechanism, recently discussed in Ref.~\cite{reina2025efficient}, involves substrate-mediated superexchange. Despite their different microscopic origins, both mechanisms lead to the same effective description, in which the exchange interaction becomes time-dependent. We capture this effect by introducing a periodically driven exchange interaction of the form
\begin{equation}
    J_f(t) = J_{f}^{(0)}+\alpha_J\,\frac{V(t)}{z_{tip}^{(0)}}
\end{equation}
where the amplitude $\alpha_J$ can be estimated from both mechanism and yields comparable values when using the parameters relevant to this system ($\alpha_J\simeq 0.1-0.6$ MHz nm$/$mV)\cite{soohion2023b,reina2025efficient}. 

In our calculations, we will also neglect the voltage dependence of the gyromagnetic factor. This approximation is well justified for small magnetic fields, where the energy shifts associated with $g$-factor modulation are expected to be small. However, at larger magnetic fields, the shifts of the energy levels scale with the external field and may become significant.  Similar considerations apply to the magnetic anisotropy terms of the Fe atom, which are not expected to play a significant role in driving spin transitions \cite{reina2025efficient}. In particular, as we show below, the modulation of the longitudinal anisotropy ${\cal D}_2$ is irrelevant in the LZSM regime \cite{rodriguez2024non, reina2025efficient}. Therefore, we neglect both $g$-factor and anisotropy modulations and focus on the dominant effects arising from the modulation of the tip magnetic field and the exchange interaction.


\subsection{Low-Energy effective Hamiltonian and avoided crossing setup }
\label{stls}


Starting from   the full Hamiltonian Eq.(\ref{htot}), in Appendix~B we derive an effective low-energy Hamiltonian   consisting of two decoupled four-dimensional blocks,   corresponding respectively to  $m_{\rm Fe}=\pm 2$. Each subspace is spanned by   the  product states of the titanium spin for a given value of $m_{\rm Fe}$, {\it {i.e.}} \{$|\uparrow, \uparrow, m_{\rm Fe}\rangle$, $|\uparrow, \downarrow, m_{\rm Fe}\rangle$, $|\downarrow, \uparrow, m_{\rm Fe}\rangle$, $|\downarrow, \downarrow, m_{\rm Fe}\rangle \}$.

Since we can set the angle of the magnetic axis near the $z$-axis, the $x$-components of the corresponding magnetic fields, $b_x$ and $B_x$, are rather small.


In order to obtain the two lowest energy levels, we diagonalize the $S_z=0$ sub-block of the Ti dimer, obtaining rotated states within this subspace. Different regimes can be identified in termos of the rotation angle $\alpha$ is given by: 

\begin{equation}
\tan{\alpha}=\frac{J_t}{x_Z},   
\end{equation}
\noindent with $x_Z = g \mu_B b_z - J_f m_{\rm Fe}$.

For the weak coupling limit, $J_t \ll |x_Z|$,  the two lowest energy states are essentially the product states $|\uparrow, \downarrow, m_{\rm Fe}\rangle$ and $|\downarrow, \uparrow, m_{\rm Fe}\rangle$, respectively. For $x_Z > 0$, the ground state  is $|\downarrow, \uparrow, m_{\rm Fe}\rangle$, whereas for $x_Z < 0$,  is $|\uparrow, \downarrow, m_{\rm Fe}\rangle$.

On the other hand, in the strong coupling regime $J_t \gg |x_Z|$, the system is naturally described in the coupled total-spin basis, where the two lowest-energy eigenstates correspond to the singlet $|S\rangle=|0,0,m_{\rm Fe}\rangle$ and triplet $|T_-\rangle=|1,-1,m_{\rm Fe}\rangle$ states of the Ti dimer, respectively. In this case, the first excited state is the singlet $S$, and the sign of $x_Z$ is less relevant than in the weak-coupling case.

Depending on the ratio $J_t/x_Z$, two different control regimes can be identified. For strong coupling, the dynamics mainly involves transitions between the singlet $S$ and the $T_-$ triplet state. In contrast, in the weak-coupling regime, control can be achieved through a spin flip of one of the Ti atoms, mediated by the Fe atom.


Restricted to the two lowest-energy level 
subspace, the Hamiltonian is given by : 
\begin{equation}\label{htls-g}
H_{\rm TLS}^{\pm} =
\left(
\begin{array}{cc}
  \frac{J_{t}}{2} -\frac{\Gamma}{2}  & \sqrt{2}\Delta_b  B_{-1,-} \\ 
\sqrt{2}\Delta_b  B_{-1,-} &  -\sqrt{2}\Delta_b( b_{z} + 2B_z) + J_{t} \pm J_f 
\end{array}
\right)
\end{equation}

\noindent where the sign depends on the projection of the Fe spin, $\Gamma = \sqrt{x^2_Z + J^2_t}$, $\Delta_b = \frac{g \mu_B}{2\sqrt{2}}$ and $B_{-1,-}$ is a rotated magnetic field given by: 

\begin{eqnarray}
B_{-1,-} = \cos\left(\frac{\alpha}{2}\right)B_x - (b_x + B_x)\sin\left(\frac{\alpha}{2}\right). \label{b-1-}   
\end{eqnarray}

As follows from the Hamiltonian  Eq.~(\ref{htls-g}), the magnetic field $B_z$ can be tuned so that the avoided crossing occurs at $V_{\mathrm{dc}}=0$, {\it i.e.}:

\begin{eqnarray}\label{bcross}
B_z = -\frac{b^{(0)}_z}{2} + \frac{J_t + 2 J^{(0)}_f + \Gamma}{4\sqrt{2}\Delta_b}.
\end{eqnarray}

\noindent With this chosen magnetic field, we can write the two-level Hamiltonian in a standard notation \cite{nori1}:

\begin{equation}\label{tlsrot}
 H_{\rm TLS} = \frac{\varepsilon}{2} \sigma_z + \frac{\Delta}{2} \sigma_x.     
\end{equation}

\noindent Here, $\varepsilon$ and $\Delta$ are given by the following expressions (see Appendix B):

\begin{eqnarray}
 \varepsilon &=& -\frac{\left(J_t + \Gamma\right)}{2} + \sqrt{2}\Delta_b b_z + 2\sqrt{2}\Delta_b B_z \pm J_f, \label{eps2level}\\
 \Delta &=& 2\sqrt{2} \Delta_b B_{-1,-}. \label{gap-1}
\end{eqnarray}



We consider operation around an equilibrium point at the avoided crossing, corresponding to $V_{\mathrm{dc}}=0$. We assume voltage-induced atomic displacements are small enough so that the tip field and $J_f$ can be accurately described by a first-order Taylor expansion about their equilibrium positions. As defined in the previous section, this gives $J_f(t)=J_f^{(0)}+\alpha_J V(t)$ and $b_z(t)=b_z^{(0)}+\alpha_z V(t)$. Retaining only terms linear in $V(t)$, we obtain:


\begin{equation}
\begin{aligned}\label{epsilonV}
\varepsilon \simeq
\Bigg[
&\left(1-\frac{x_Z^{(0)}}{\Gamma^{(0)}}\right)
\sqrt{2}\Delta_b \alpha_z \\
&\pm\left(1+\frac{x_Z^{(0)}}{\Gamma^{(0)}}\right)\alpha_J
\Bigg] V(t).
\end{aligned}
\end{equation}

\noindent where $x_Z^{(0)}$ and $\Gamma^{(0)}$ denote the equilibrium values of $x_Z$ and $\Gamma$, respectively, at $V_{\mathrm{dc}}=0$.





The above equations fully determine the low-energy effective dynamics and provide a simple framework for describing quantum-state manipulation around the avoided crossing. The Ti$_2$ dimer without the nearby Fe atom is naturally recovered by setting $J_f=0$ and $\alpha_J=0$ in the expressions above.

\section{Engineering LZSM Interference in the Titanium Dimer}
\label{stidim}

In this section, we show how to engineer avoided level crossings in a Ti dimer directly in the laboratory frame, without relying on resonant driving. This minimal, experimentally accessible system is an excellent platform to identify how to open and control the relevant gaps. Our approach exploits two experimentally tunable ingredients: the interatomic exchange coupling, set by the Ti–Ti distance, and the orientation and magnitude of the external magnetic field.



\begin{figure}[hbt]
\includegraphics[width=0.95\linewidth]{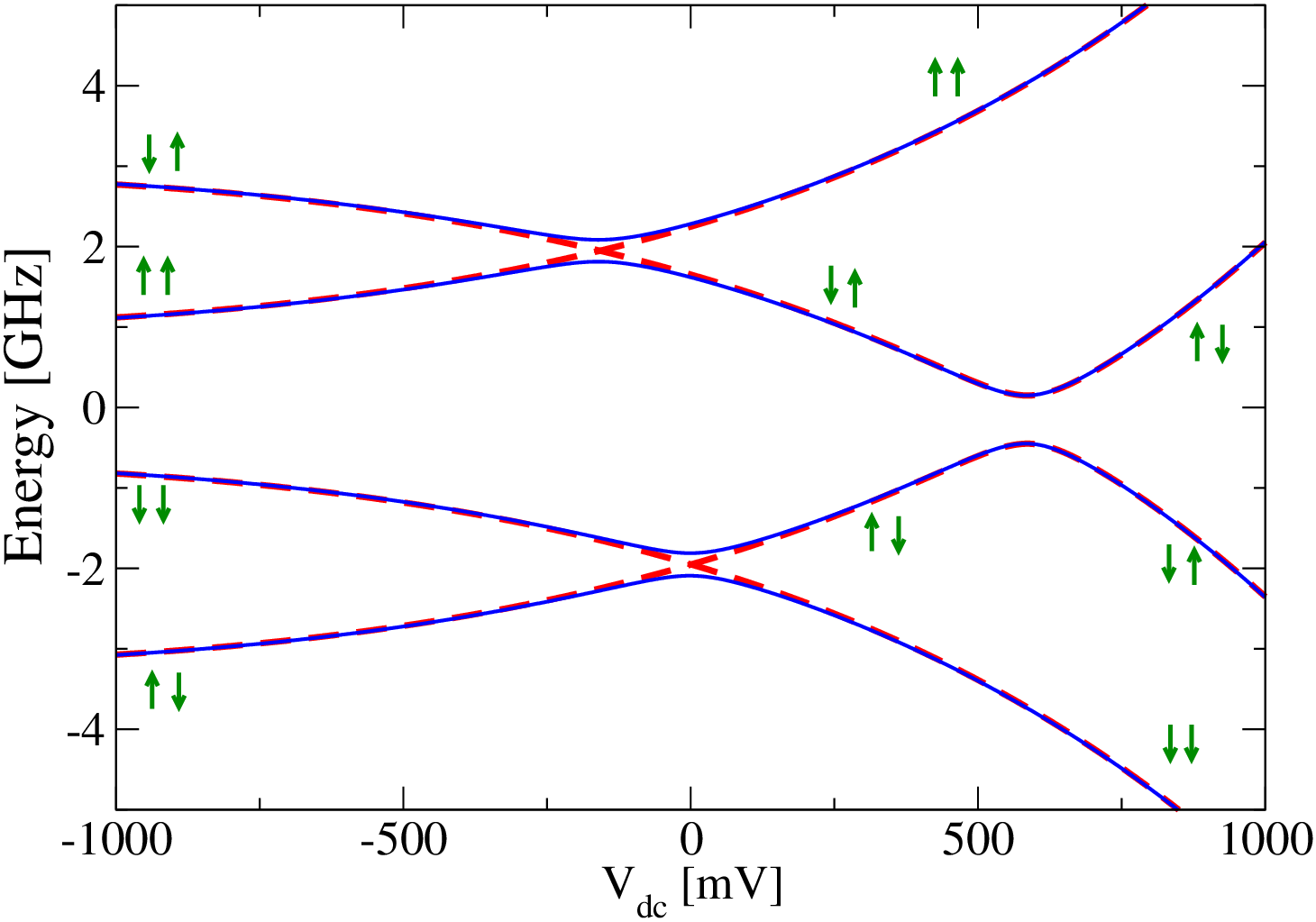}
\caption{\label{figAv1} Energy spectrum of the Ti dimer obtained from exact diagonalization of Eq. (\ref{htot}),  $d_{\mathrm{tip}}=0.525$ nm ($b_z<0$) , $B_{\mathrm{ext}}=0.1385$~T, $J_t=0.6$~GHz. Red dashed: $\theta=90^\circ$; blue: $\theta=89^\circ$. The vertical green arrows indicate the asymptotic states of the first and second spins far from the avoided crossing. 
}
\end{figure}

Figure~\ref{figAv1} shows the spectrum of a Ti dimer with $J_t=0.6$ GHz \cite{kot2023electric}, without the nearby
Fe atom and with an isotropic tip located above the first Ti atom. Throughout this section, we focus on
the weak-coupling regime, choosing tip--atom distances such that $J_t<g\mu_B|b_z|$. For smaller interatomic
separations, coupling strengths as large as $60$ GHz can be reached \cite{kot2023electric}. The spectrum
is plotted as a function of the DC voltage applied through the tip, $V_{\mathrm{dc}}$. We denote the ground
and first excited states by $|\psi_0\rangle$ and $|\psi_1\rangle$, respectively. For a perpendicular
external magnetic field, their energies cross at $V_{\mathrm{dc}}=0$. The magnetic field at which the crossing occurs can
be obtained from Eq.~\eqref{bcross} evaluated at $J_f=0$. For a coupling of $0.6$ GHz, as in the figure,
the required external magnetic field is approximately $130$ mT, whereas for strongly coupled Ti dimers,
with $J_t$ close to $60$ GHz, it exceeds $2$ T.

When the magnetic field is fully out of plane (red-dashed curves), the ground and first excited states cross without opening a gap. In contrast, an avoided crossing associated with the Ti--Ti exchange coupling appears between the first
and second excited states, $|\psi_1\rangle$ and $|\psi_2\rangle$, at larger DC voltages, as observed experimentally in recent ESR-STM studies \cite{kot2023electric,wang2025electrically}. Introducing a small in-plane magnetic-field component (blue curves, $\theta=89^\circ$) opens additional gaps at the other crossings. The absence of such gaps in the experiment of Kot \textit{et al.}~\cite{kot2023electric} is consistent with an almost fully out-of-plane field. However, in-plane components induced by tip anisotropy may also contribute to opening a gap.

The sign of $b_z$, determined by the balance between the dipolar and exchange contributions, selects the mechanism coupling the two relevant levels and determines their spin character. Depending on the tip position, the avoided crossing involves the diabatic configuration $|\downarrow\downarrow\rangle$ and either $|\uparrow\downarrow\rangle$ or $|\downarrow\uparrow\rangle$, corresponding to a flip of the Ti spin beneath the tip or of the remote Ti spin, respectively.

When the tip field is dominated by the exchange contribution and $b_z>0$, the energy gap $\Delta$ between the ground and first excited states can be obtained from Eq.~(\ref{gap-1}), in the weak-coupling limit, by expanding in powers of $J_t/b_z^{(0)}$,

\begin{equation} \label{deltabz+}
\Delta \simeq -\sqrt{2}\Delta_b B_x
+\frac{\Delta_b J_t}{g\mu_B b_z^{(0)}}\left(B_x-b_x\right).
\end{equation}

\noindent Thus, for small $J_t$, the leading contribution to the gap depends only on the external in-plane magnetic field $B_x$, while the tip-induced contribution enters only as a correction proportional to $J_t/b_z^{(0)}$. In this regime, far from the avoided crossing, the ground and first excited states, $|\psi_0\rangle$ and $|\psi_1\rangle$, approximately coincide with the diabatic configurations $|\downarrow\downarrow\rangle$ and $|\downarrow\uparrow\rangle$, with their energy ordering depending on the side of the crossing. Transitions between these configurations would therefore allow the remote Ti spin to be flipped while leaving the spin beneath the tip unchanged. However, the modulation of the detuning is also proportional to $J_t/b_z^{(0)}$,
as shown in Eq.~\eqref{epsilonV} for $x_z>0$ and in the absence of the Fe atom. For small $J_t$, controlling the spin that is not located directly below the STM tip, the remote spin, is expected to become highly inefficient.


For $b_z<0$, corresponding to the regime in which the dipolar contribution dominates the tip field, the situation changes and the system becomes controllable (see Eq.~(\ref{epsilonV}) for $x_Z^{(0)}<0$, neglecting the terms associated with the Fe atom). In this regime, the gap-opening mechanism can be understood from Eq.~(\ref{gap-1}) for $x_Z<0$ (see Appendix~\ref{Asc} for further details), giving the gap


\begin{equation}
\Delta \simeq  - \Delta_b (B_x+b_x) - \Delta_b B_x\left( \frac{ J_t}{g \mu_B b^{(0)}_z} \right) \label{deltabz-},
\end{equation}


\noindent which depends on both the external in-plane field and the anisotropy-induced in-plane tip field.
In this regime, far from the avoided crossing, the ground and first excited states, $|\psi_0\rangle$ and
$|\psi_1\rangle$, approximately coincide with $|\downarrow\downarrow\rangle$ and $|\uparrow\downarrow\rangle$, with their energy ordering depending on the side of the crossing. Transitions between these diabatic configurations flip the Ti spin
directly beneath the tip, for which the electrical modulation provides more efficient control.

The dependence of the gap on the transverse tip field also provides insight into the role of tip anisotropy in ESR-STM experiments. In the experiment by Kot \textit{et al.}, the absence of additional avoided crossings suggests that the effect of the transverse field generated by the tip is not large enough to open experimentally resolvable gaps. This could be consistent with either a very small tip anisotropy or a tip operating in a regime dominated by exchange.

\begin{figure*}
    \centering
    \includegraphics[width=0.8\linewidth]{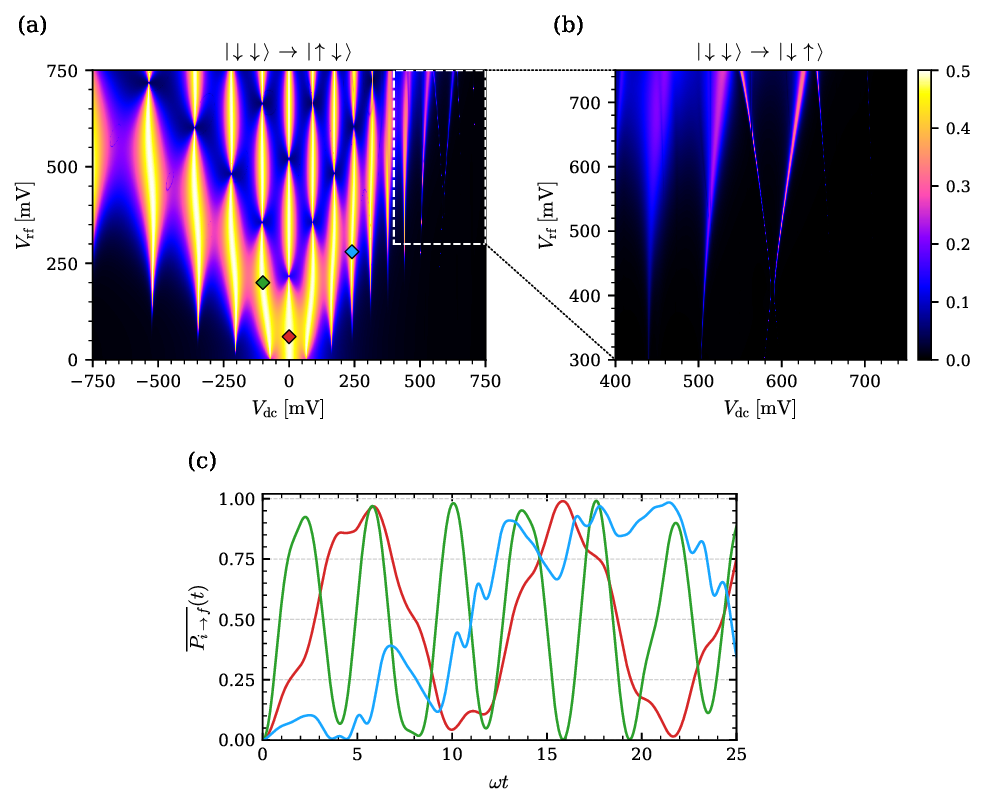}
\caption{\label{figLZSMwea} Contour plot of LZSM interference for coupled Ti atoms with $J_t = 0.6$ GHz. The STM tip placed at $d_{tip}=0.525$ nm ($b_z<0$) and has no anisotropy. The driving frequency is $\omega=0.4$ GHz. The external magnetic field is chosen such that the system becomes degenerate when there are no in-plane components ($B=0.1385$ T), and its angle is set to $\theta=89^\circ$ to open a small gap. (a) Averaged transition probability map between the state $|\downarrow\downarrow\rangle$ and the state $|\uparrow\downarrow\rangle$. (b) Averaged transition probability map between the state $|\downarrow\downarrow\rangle$ and the state $|\downarrow\uparrow\rangle$. (c) Transition probability between the state $|\downarrow\downarrow\rangle$ and the state $|\uparrow\downarrow\rangle$ as a function of time for the voltage values indicated in the upper panel. }
\end{figure*}

Assuming that the system is prepared in $|i\rangle=|\downarrow\downarrow\rangle$ at an initial time $t_0$, the time-averaged transition probability to $|f\rangle$ is given by


\begin{equation}
    \overline{\overline{P_{if}}} = \lim_{t' \to \infty} \frac{1}{t'} \int_0^{t'} dt \, \overline{P}_{if}(t)  ,
\end{equation}
where
\begin{equation}
    \overline{P_{if}}(t) = \frac{1}{T} \int_0^T  dt_0 \, \left|\langle f|\psi(t,t_0)\rangle\right|^2  ,
\end{equation}

\noindent and $|\psi(t,t_0)\rangle = \hat{U}(t_0+t,t_0)\,|i\rangle$ is the state obtained by evolving $|i\rangle$ from $t_0$ during a time $t$, and $T=2\pi/\omega$ is the period of the driving field. The double bar indicates a double average: over the initial time $t_0$ within one driving period, which accounts for the unknown phase of the drive at the moment the driving starts, and over the elapsed time $t$ in the long-time limit. The calculations were performed within the Floquet formalism \cite{ferron2010} for the full four-level Hamiltonian of the Ti dimer, and the results are plotted in Fig.~\ref{figLZSMwea}.

In Fig.~\ref{figLZSMwea} (a), the LZSM map of $\overline{\overline{P}}_{i\rightarrow f}$ for $|f\rangle=|\uparrow\downarrow\rangle$   as a function of $V_{\mathrm{dc}}$ and $V_{\mathrm{rf}}$ is plotted for a Ti dimer with $J_t=0.6$ GHz, using the same parameters as in Fig.~\ref{figAv1} and a driving frequency $\omega=400$ MHz. The isotropic STM tip is located above the first Ti atom at a height of $0.525$ nm (dipolar tip, $b_z<0$). 


The first notable feature is that  for specific values of the DC and RF applied voltages, nearly complete population transfer, corresponding to values of $\overline{\overline{P}}_{i\rightarrow f}\sim 0.5$, is achieved. This reflects an effective flip of the Ti spin located beneath the tip, enabling coherent control of the dimer through a LZSM protocol. A second important feature of the LZSM map is its pronounced asymmetry with respect to $V_{\mathrm{dc}}$, that is mainly due to the asymmetry of the energy spectrum shown in Fig. \ref{figAv1}.

In addition, for sufficiently large positive values of $V_{\mathrm{dc}}$ and/or $V_{\mathrm{rf}}$, the driving protocol can reach the second and third avoided crossings, enabling transitions involving the higher-energy eigenstates $|\psi_2\rangle$ and $|\psi_3\rangle$. These eigenstates have appreciable weight in the states $|\downarrow\uparrow\rangle$ and $|\uparrow\uparrow\rangle$, whose populations were
negligible in the two-level regime. Their participation qualitatively modifies the interference pattern and resonance structure. To analyze this, in Fig.~\ref{figLZSMwea} (b) we plot $\overline{\overline{P}}_{i\to f}$ with
$|f\rangle = |{\downarrow\uparrow}\rangle$, i.e., the probability of flipping the remote spin by driving through the second avoided crossing. As can be seen, this process is possible to a modest extent, but it does not appear to be efficient, at least in the present configuration.




Finally, the lower panel of Fig.~\ref{figLZSMwea} shows the transition probability as a function of time for selected pairs $(V_{\mathrm{dc}}, V_{\mathrm{rf}})$ (blue, red, and green squares in the upper panel). In other words, for fixed voltage parameters, we track the population of the target state over time. For the representative points considered, we find that transition probabilities as high as $98\%$ can be achieved on very short timescales, well below the experimentally reported coherence times, which are close to $100$~ns \cite{wang2023qb,wang2023universal}. Considering the first peaks, the red and blue curves reach transition probabilities of $98.9\%$ and $98.4\%$ at $t\simeq39.6$ and $53.7$~ns, respectively. The green curve reaches $98.1\%$ at $25.0$~ns and $99.1\%$ at $44.0$~ns.


In summary, the characteristics and efficiency of LZSM control depend on the mechanism that opens the avoided crossing. When $b_z>0$, the gap can be tuned through experimentally controllable parameters, and the protocol addresses the remote Ti spin. However, remote control remains inefficient in this configuration. When the tip is retracted enough to have $b_z<0$, control is efficient but addresses the Ti spin directly beneath the tip, and the gap depends on the tip characteristics.

Accessing higher avoided crossings offers alternative pathways for remote spin control. Our calculations show that reaching the second avoided crossing allows us to flip the remote Ti spin. Nevertheless, efficient remote control requires new mechanisms that can be provided by a nearby Fe atom, as discussed in the next section.


\section{Remote quantum control}
\label{sremote}

Achieving remote spin control—that is, manipulating a spin not directly beneath the STM tip—is a central goal in this class of experiments, as it would protect the controlled spin from direct exposure to the tunneling electrons injected by the STM tip. As shown in the previous section, however, such control is extremely challenging in an atomic architecture composed solely of two Ti atoms. Introducing an Fe atom not only modifies the energy spectrum but also provides an additional mechanism to control spin dynamics. When the tip is positioned at distances comparable to those considered in the previous section, the modulation of the tip--Ti coupling and the Ti--Fe interaction compete. As the tip is retracted from the surface, the modulation of the coupling between the tip and the first Ti atom becomes progressively less relevant, and the modulation of the Ti--Fe interaction becomes the dominant control mechanism \cite{soohion2023b,phark2023double,wang2023qb}.

\begin{figure}[hbt]
\centering
\includegraphics[width=0.9\linewidth]{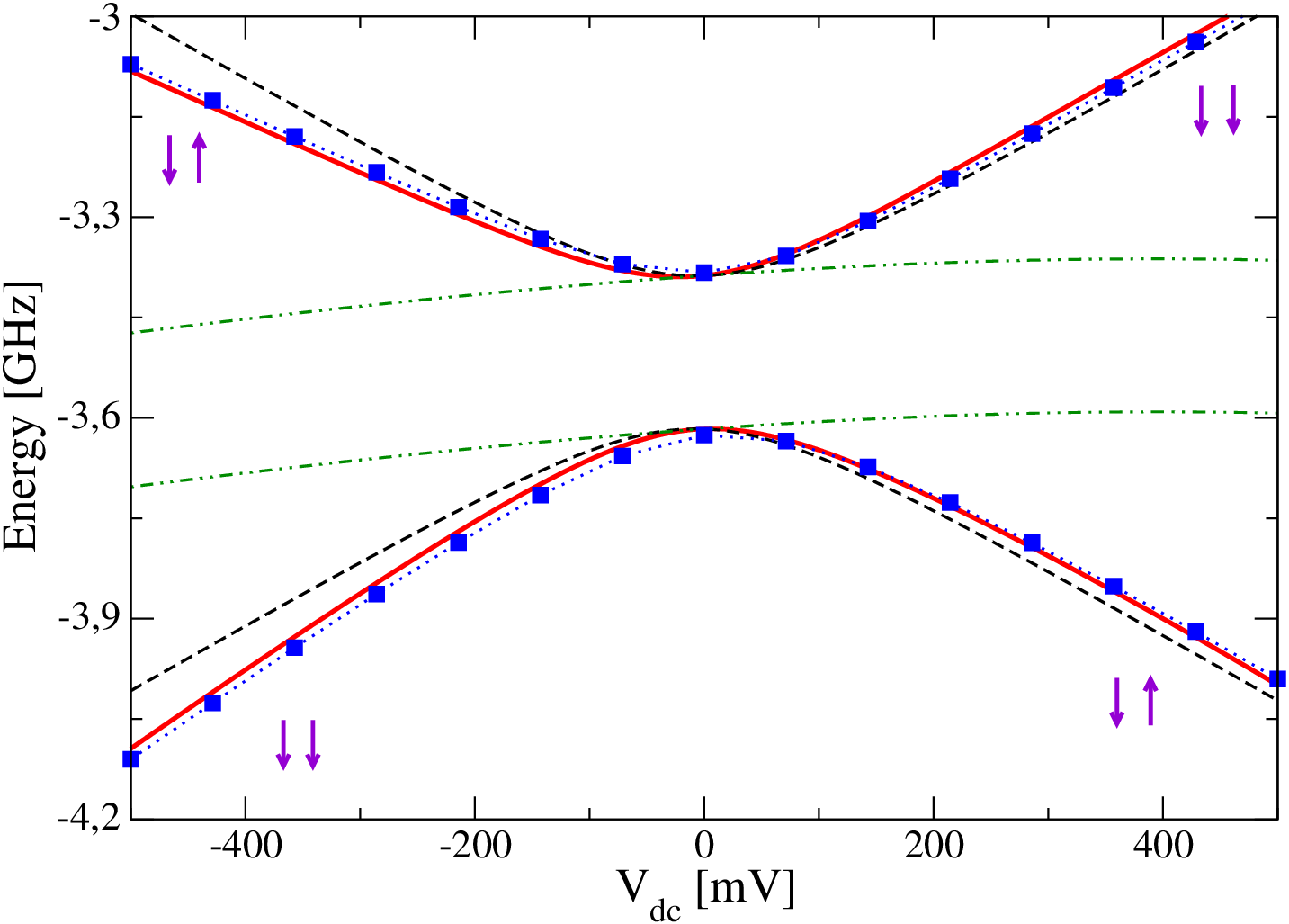}
\caption{\label{figAv2} Energy spectrum of the Ti-Ti-Fe trimer obtained from exact diagonalization of Eq. (\ref{htot}) (solid red line) and from the two-level approximation using Eqs.~\eqref{htls-g} (blue squares). Black dashed lines show the avoided crossing when the modulation of the tip-atom distance is not present, and green dashed lines show the same in the case that the modulation of the Ti-Fe coupling is not considered (in both cases, the calculations were performed by exact diagonalization of the complete system). All the calculations were performed for $J_f=6.8$~GHz, $\theta=89^\circ$, $d_{\mathrm{tip}}=0.6$ nm, $B_{\mathrm{ext}}=0.497$~T, and $J_t=0.6$~GHz.}   
\end{figure}

As we have already mentioned, by tuning the external magnetic field the system can be brought to a regime where the two lowest-energy levels cross at $V_{\mathrm{dc}}=0$, with the field strength estimated using Eq.~(\ref{bcross}). As in the dimer case, these crossings can be lifted, opening a gap. The inclusion of the Fe atom, however, makes the situation more complex, as three energy scales must now be considered: the Ti--Ti coupling, the Fe--Ti coupling involving the remote Ti atom, and the coupling between the STM tip and the Ti atom directly beneath it. We focus on the regime in which the remote Ti spin is controlled through the Fe--Ti interaction. For this condition to be satisfied, the tip must be retracted sufficiently far from the surface so that $2J_f>|g\mu_B b_z|$, and then $2J_f+g\mu_B b_z >0$. In this regime, when $J_t\ll 2J_f+g\mu_B b_z$, Eq.~(\ref{gap-1}) shows that the gap is determined almost exclusively by the in-plane component of the external magnetic field.

In Fig.~\ref{figAv2}, we show the two lowest-energy levels of the dimer considered in the previous section after adding an Fe atom at a distance of $0.59$ nm from the second Ti ($J_f=6.8$ GHz) \cite{soohion2023b,wang2023qb}. In the experiments, the Fe atom can typically be placed at two symmetry-related positions with respect to the Ti dimer \cite{soohion2023b,wang2023qb}. Here, we consider one of these configurations. The Ti atoms are arranged in the same configuration as in the previous section, yielding $J_t=0.6~\mathrm{GHz}$ and therefore satisfying the condition $J_t\ll J_f$. The red solid lines correspond to the exact diagonalization of the full ($20\times20$) Hamiltonian Eq. \eqref{htot}, while the blue dots show the effective two-level approximation derived in Sec. \ref{stls} (see Eq.~(\ref{htls-g})) for the Fe spin in the state $m_{Fe}=-2$. For comparison, we also include the full calculation neglecting the modulation induced by the tip (black dashed lines), as well as the case where the modulation of the exchange coupling between the Fe atom and the second Ti atom is neglected (green dashed lines). For the parameters considered here, particularly for $d_{\mathrm{tip}}=0.6$ nm, the DC modulation induced directly by the tip is found to be practically negligible. Therefore, a plausible  driving strategy could be  the  time dependent modulation of the exchange interaction between the Fe and Ti \cite{soohion2023b,wang2023qb}, as we describe in what follows.

\begin{figure}
    \centering
    \includegraphics[width=1.\linewidth]{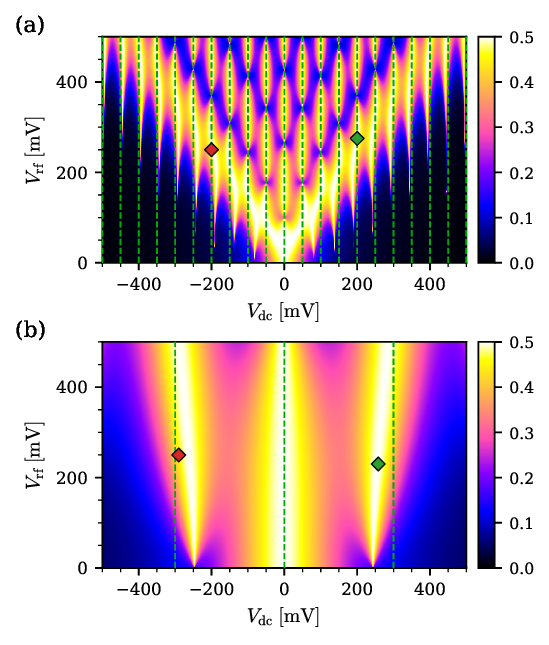}
        \caption{Time-averaged LZSM transition maps between diabatic states, $|--Fe\rangle\rightarrow|-+Fe\rangle$ for the trimer with $J_t=0.6$ GHz and $J_f=6.8$ GHz, shown for different orientations of the magnetic field and different driving frequencies. (a) $\theta=89.5^\circ$ and $\omega=0.1$ GHz. (b) $\theta=88.5^\circ$ and $\omega=0.6$ GHz. }
    \label{lzsmjf-w}
\end{figure}

Far from the avoided crossing, the purple arrows indicate the spin projections of the two Ti atoms for each energy level, providing a simple picture of the character of the corresponding states. Population can be transferred from the $|\downarrow\downarrow\rangle$ configuration to a state in which only the remote Ti spin is flipped, demonstrating that LZSM interferometry can be used for remote control of an atomic spin qubit. This is well illustrated in Fig.~\ref{lzsmjf-w}, where we show the time-averaged transition probability as a function of $V_{\mathrm{dc}}$ and $V_{\mathrm{rf}}$ for a Ti dimer with $J_t=0.6$ GHz and an Fe atom located $0.59$ nm from the second Ti atom. The results show that a high degree of remote control can be achieved, avoiding unwanted effects associated with the tunneling current. Importantly, this control is obtained using relatively low driving frequencies compared to usual ESR-STM experiments \cite{rodriguez2024non}. Figure~\ref{lzsmjf-w} compares two different parameter sets. The main differences between them are the driving frequencies and the size of the gaps, whose magnitude is determined by the orientation of the magnetic field. As shown in Eq.~(\ref{gap-1}), in the relevant regime the gap is approximately given by $\Delta \simeq g\mu_B B_x$. In the same limit, the equilibrium mixing angle satisfies
$\tan(\alpha^{(0)})\simeq 0$, so that $x_Z^{(0)}\simeq\Gamma^{(0)}$. Equation~\eqref{epsilonV} then yields

\begin{eqnarray}
     \varepsilon(t)=-2\alpha_J\left( V_{\mathrm{dc}}+V_{\mathrm{rf}}\sin{(\omega t)}\right)
\end{eqnarray}

\noindent With these expressions, the effective Hamiltonian takes the form of Eq.~\eqref{tlsrot}, with the gap set by the
transverse magnetic field and the static detuning and driving amplitude controlled by the applied voltage through the Ti-Fe exchange coupling.


\begin{figure*}
    \centering
    \includegraphics[width=0.85\linewidth]{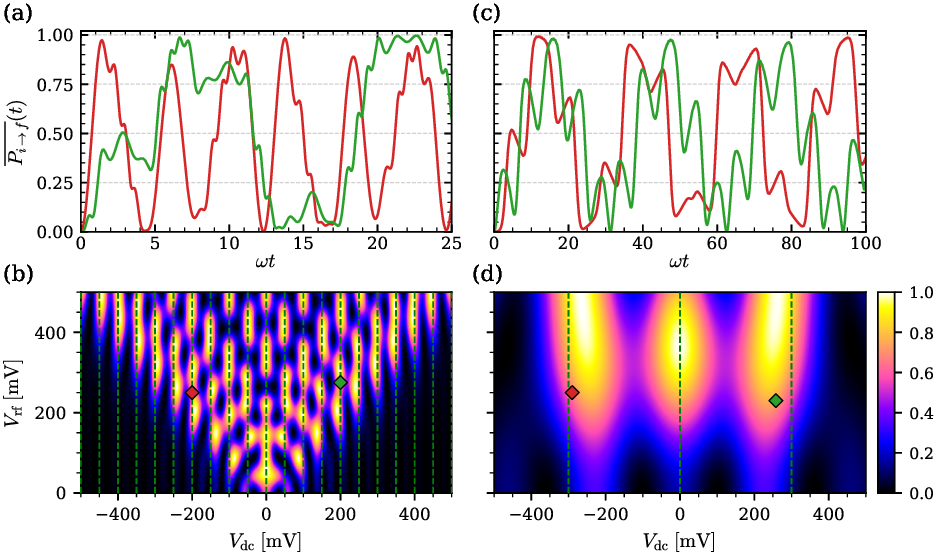}
        \caption{Time evolution LZSM transition between diabatic states, $|--Fe\rangle\rightarrow|-+Fe\rangle$ for the trimer with $J_t=0.6$ GHz and $J_f=6.8$ GHz, $d_{\rm tip}=0.6$ nm, $B_{\rm ext}=0.497$ T. (a) Time-dependent populations for the pairs of $V_{\rm dc}$ and $V_{\rm rf}$ indicated in the upper panel of Fig.~\ref{lzsmjf-w} for $\theta=89.5^\circ$, and $\omega=0.1$ GHz. (b) Map of the transition probability evaluated at an operation time satisfying $\omega t=3\pi$, $\theta=89.5^\circ$, and $\omega=0.1$ GHz. (c) Time-dependent populations for the pairs of $V_{\rm dc}$ and $V_{\rm rf}$ indicated in the upper panel of Fig.~\ref{lzsmjf-w} for $\theta=88.5^\circ$, and $\omega=0.6$ GHz. (d) Map of the transition probability evaluated at an operation time satisfying $\omega t=3\pi$, $\theta=88.5^\circ$, and $\omega=0.6$ GHz.
\label{RemoteTime-1}}
    \label{eficient}
\end{figure*}


In the fast-driving regime described in Appendix~\ref{Alzsm}, where the adiabaticity parameter $\delta=(\mu_B g B_x)^2/(8\hbar\alpha_J V_{\mathrm{rf}}\omega)$ satisfies $\delta\ll1$ \cite{nori1,nori2}, the probability of selectively flipping the remote Ti spin while leaving the Ti spin beneath the tip unchanged can be estimated as

\begin{eqnarray}\label{fastdriving}
P_{remote}(t)&=&\sum_n \frac{\Gamma_n^2}{2\Omega_n^2}\left(1-\cos{(\Omega_n t)}\right)
\end{eqnarray}

\noindent where $P_{remote} \equiv P_{|\downarrow,\downarrow,m_{\mathrm{Fe}}=-2\rangle\rightarrow|\downarrow,\uparrow,m_{\mathrm{Fe}}=-2\rangle}$, 
$\Omega_n^2=\left(n\omega-2\alpha_J V_{\mathrm{dc}}\right)^2+\Gamma_n^2$, $\Gamma_n=(-1)^n\Delta J_n(2\alpha_J V_{\mathrm{rf}}/\omega)$, and $J_n$ denotes the Bessel function of the first kind.




\noindent At a given $n$-photon resonance, Eq.~\eqref{fastdriving} predicts an oscillation period $2\pi/\Gamma_n\propto 1/\Delta$. Since the gap $\Delta$ is mainly controlled by the in-plane component of the external magnetic field, $B_x$, increasing this component can shorten the time required to flip the remote Ti spin. This tuning is constrained by the validity of Eq.~\eqref{fastdriving}, so the system must remain in the fast-driving regime, and the resonances must be sufficiently well separated. 

To assess whether the fast-driving condition is satisfied, we evaluate the adiabaticity parameter $\delta$ for the parameters used in Fig.~\ref{lzsmjf-w}. As the driving amplitude increases from $V_{\mathrm{rf}}=100$ to $500$ mV, $\delta$ decreases from about $0.3$ to $0.06$ in the upper panel and from $0.5$ to $0.1$ in the lower panel. Thus, the condition $\delta\ll 1$ is better satisfied at the larger driving amplitudes considered here.



The dashed green lines in Fig. \ref{lzsmjf-w} indicate the resonance positions predicted by Eq.~(\ref{fastdriving}). The comparison with the numerical results is consistent with the values of $\delta$: the fast-driving approximation requires $\delta\ll 1$, which is not fully satisfied in most of the parameter range considered here. As shown in Fig. \ref{lzsmjf-w}, for the small-gap case (upper panel), the theoretical lines agree well with the numerical resonances at large driving amplitudes, where, according to the adiabaticity parameter, the system is close to the fast-driving regime. In the large-gap case (lower panel), where larger driving frequencies are used, the fast-driving regime is more difficult to reach. Nevertheless, it is interesting to note that, at large driving amplitudes, the theoretical lines start to approach the numerical resonances.

This leads to a trade-off between increasing the gap and preserving the validity of the fast-driving regime. Although the gap can be enlarged by increasing the in-plane component of the magnetic field, this also requires higher driving frequencies and larger driving amplitudes to remain in the fast-driving regime and, consequently, larger voltages to satisfy the resonance conditions.

In Figs.~\ref{RemoteTime-1} (a) and (c), we show the probabilities of flipping the remote spin as a function of time for the two cases discussed in Fig.~\ref{lzsmjf-w} and for two selected pairs of $V_{\mathrm{dc}}$ and $V_{\mathrm{rf}}$, corresponding to the green and red points indicated in the two panels of Fig.~\ref{lzsmjf-w}. These points were chosen at the fourth (upper panel), and first (lower panel) resonances, where the time-averaged transition probability displayed in the LZSM maps attains a maximum value, indicating that remote control should be possible. The time evolution confirms this expectation: the remote Ti spin can be inverted with nearly unit probability on a nanosecond timescale. 

Figure~\ref{RemoteTime-1}(a) shows that the red curve reaches a spin-inversion probability of $97.3\%$ at $14.2$~ns and $98.3\%$ at $137.5$~ns, while the green curve reaches $98.7\%$ at $66.6$~ns. In Fig.~\ref{RemoteTime-1}(c), the red and green curves reach inversion probabilities of $99.3\%$ and $98.1\%$ at approximately $19.8$ and $26.5$~ns, respectively. These operation times are well below the experimentally measured coherence times of the remote spin $T_2\sim300$ ns \cite{wang2023qb}.  Of course, these results depend on the chosen driving amplitudes and detuning values. The lower panels of Fig.~\ref{RemoteTime-1} show the transition probability at a fixed time, chosen such that $\omega t=3\pi$. These maps reveal extended regions in which the remote qubit can be efficiently controlled on short time scales. In the small-gap case, where the fast-driving regime is easier to reach, these regions are more directly connected to the resonance conditions and can be identified by following the resonance lines while avoiding the zeros of the Bessel functions. For a larger gap, where the driving frequency is increased, but the system is slightly farther from the fast-driving regime, the optimal regions still appear close to the resonances. Although the agreement with the fast-driving prediction is less precise than in the previous case, the regions of good performance are broader and less sharply defined, providing slightly more flexibility in choosing the operation voltages.


\section{Conclusions}
\label{sc}


 We propose a protocol for the non-resonant electric control of atomic spin qubits based on Landau-Zener-Stückelberg-Majorana (LZSM) interferometry. As a minimal platform, we analyze Ti dimers on MgO, where avoided level crossings can be  engineered in the laboratory frame. We show that the Ti–Ti exchange interaction, in combination with the orientation of the external magnetic field and the local field generated by the STM tip, provides a flexible route of tuning the relevant energy gaps.

The central result of this work is the coherent remote spin control in a Ti--Ti--Fe trimer using LZSM interferometry. With the STM tip positioned above one Ti atom, the protocol selectively flips the other Ti spin while leaving the spin beneath the tip unchanged. The Fe atom enables this control by providing an electrically tunable exchange coupling to the remote Ti atom. Modulation of this coupling drives the system through an avoided crossing, allowing the remote spin to be controlled even when the tip is retracted sufficiently to suppress tip-induced magnetic-field modulation. This geometry separates the controlled qubit from the tunneling current, offering a route to mitigate tunneling-induced decoherence in STM-based spin manipulation.

A particularly relevant feature of the atomic engineered system is that, in some conditions, the size of the avoided-crossing gap becomes controlled almost exclusively by the in-plane component of the external magnetic field, $\Delta \simeq g\mu_B B_x$. This is experimentally important because it removes the dependence of the gap on poorly controlled microscopic details, such as the STM-tip anisotropy. Importantly, there are experimentally accessible conditions under which the operation time can be tuned by adjusting the in-plane component of the external magnetic field. Our numerical simulations show that high-probability remote spin flips can be achieved on nanosecond time scales. In particular, we obtain good population inversion for operation times of the order of $15$ ns.

Overall, our results show that electrically driven LZSM interferometry provides a viable route for fast, non-local quantum control in engineered atomic spin structures. The Ti-Ti system illustrates how avoided crossings can be designed, while the Ti-Ti-Fe geometry demonstrates that these crossings can be exploited for remote qubit manipulation. This establishes remote LZSM-STM as a promising strategy for controlling atomic qubits with reduced sensitivity to tunneling current and with experimentally tunable operation times.

\section*{Acknowledgments}

We are grateful to Joaquín Fernadez-Rossier, Kai Yang and Christian Ast for fruitful discussions and for critically reading the manuscript. AF and SSG acknowledge funding from CONICET (PIP11220200100170 and PIP11220250100045CO) and UNNE. AF, MJS and DD acknowledge partial financial support from ANPCyT (PICT2019-0654). MJS, DD and GMMA  acknowledge support from CNEA, UNCuyo and CONICET (PIP 11220220100212CO).

\newpage

\newpage

\clearpage
\onecolumngrid
 
\section*{Appendix}

\appendix

\renewcommand{\thefigure}{A\arabic{figure}}
\setcounter{figure}{0}

\section{Two Level Hamiltonian for the Ti-Ti-Fe}\label{Asc}

Since the magnetic field is oriented nearly perpendicular to the substrate surface, with an angle close to $89^{\circ}$, the energy contributions associated with $g\mu_B b_x$  and $g\mu_B B_x$ are much smaller than the other energy scales in the Hamiltonian. Also, we assume negligible tip anisotropy. Under this assumptions, we can work in the laboratory frame, with the $z$ axis chosen perpendicular to the substrate surface. In addition, Fe on MgO exhibits strong magnetic anisotropy, so that the spin doublet with $m_{Fe}=\pm2$ is well separated from the remaining $S=2$ states ($m_{Fe}=-1,0,+1$). Although the two states of this doublet are weakly coupled \cite{rodriguez2024non}, the Zeeman splitting is much larger than this coupling. Therefore, to first approximation, the coupling can be neglected. As a result, the Fe states can be treated as effectively decoupled, and the low-energy Hilbert space of the Ti-Ti-Fe system separates into two Ti-Ti blocks, each experiencing an effective magnetic field generated by the Fe atom.

We start from the Hamiltonian in Eq.~\eqref{htot}, written in the uncoupled Ti-Ti basis ${|\uparrow,\uparrow\rangle, |\uparrow,\downarrow\rangle, |\downarrow,\uparrow\rangle, |\downarrow,\downarrow\rangle}$, with the $z$ axis taken perpendicular to the substrate surface and for a fixed Fe spin projection, $m_{\mathrm{Fe}}$:

\begin{equation}\label{hrot2_reordered_1423}
H_{t} =
\left(
\begin{array}{cccc}
J_{t} + J_f \frac{m_{\rm Fe}}{2} +\sqrt{2}\Delta_b (b_{z} +2B_z) & \sqrt{2}\Delta_b  B_x & \sqrt{2}\Delta_b (b_x + B_x) & 0 \\ 
 \sqrt{2}\Delta_b  B_x & \frac{J_{t}}{2} - J_f \frac{m_{\rm Fe}}{2} +\sqrt{2}\Delta_b b_{z} & \frac{J_t}{2} & \sqrt{2}\Delta_b (b_x + B_x) \\ 
 \sqrt{2}\Delta_b (b_x + B_x) & \frac{J_t}{2} & \frac{J_{t}}{2} + J_f  \frac{m_{\rm Fe}}{2} -\sqrt{2}\Delta_b b_{z} & \sqrt{2}\Delta_b  B_x \\ 
 0 & \sqrt{2}\Delta_b (b_x + B_x) & \sqrt{2}\Delta_b  B_x &  -\sqrt{2}\Delta_b( b_{z} + 2B_z) + J_{t} - J_f \frac{m_{\rm Fe}}{2}
\end{array}
\right)
\end{equation}

\noindent where $\Delta_b=\frac{g \mu_B}{2\sqrt{2}}$.

If we temporarily neglect $b_x$ and $B_x$, which are much smaller than their corresponding $z$ components, the Hamiltonian is diagonal except for the terms coupling the states within the $S^{(z)}=0$ (where $S^{(z)}$ is the total spin of the two Ti atoms of the dimer) subspace. In this subspace, we have:

\begin{equation}\label{hrot2_reordered_1423}
H_{Sz=0} =
\left(
\begin{array}{cc}
  \frac{J_{t}}{2} - J_f \frac{m_{\rm Fe}}{2} +\sqrt{2}\Delta_b b_{z} & \frac{J_t}{2}  \\ 
  \frac{J_t}{2} & \frac{J_{t}}{2} + J_f  \frac{m_{\rm Fe}}{2} -\sqrt{2}\Delta_b b_{z}  \\ 
\end{array}
\right)
\end{equation}

It is convenient to transform the Hamiltonian into a basis in which it becomes diagonal when the in-plane components of the magnetic fields are neglected. This can be achieved by performing a rotation within the $S^{(z)}=0$ subspace.

\begin{eqnarray}
|\Psi_{+,\alpha} \rangle = \cos{\frac{\alpha}{2}} | \uparrow \downarrow \rangle -  \sin{\frac{\alpha}{2}} | \downarrow \uparrow \rangle ,\\ 
|\Psi_{-,\alpha} \rangle = \cos{\frac{\alpha}{2}} | \downarrow \uparrow  \rangle + \sin{\frac{\alpha}{2}} | \uparrow \downarrow \rangle ,
\end{eqnarray} 

\noindent with 

\begin{eqnarray}\label{xz}
&\tan{(\alpha)} = \frac{J_t}{x_Z}, \\ 
&x_Z = 2 \left( \sqrt{2}\Delta_b b_z -J_f \frac{m_{Fe}}{2}\right),  
\end{eqnarray}

\noindent and the corresponding energies are given by

\begin{eqnarray}
E\pm = \frac{J_t}{2} \pm \frac{\Gamma}{2}, \\
\Gamma = \sqrt{ x_Z^2 + J_t^2}. 
\end{eqnarray}

\noindent The angle $\alpha$ is determined by the ratio $J_t/x_Z$. While $J_t$ is always positive, $x_Z$ may be either positive or negative. The sign of the contribution associated with $J_f$ depends on the spin projection of the Fe atom.  On the other hand, the $z$ component of the tip magnetic field, given in Eq.~\eqref{b0z}, can be written as

\begin{equation}
b_z=(b_{\rm exc}-b_{\rm dip})\cos(\theta+\eta).
\end{equation}

\noindent The tip contribution vanishes at the so-called NOTIN position, where the exchange and dipolar fields exactly cancel. For tip positions farther from the sample than the NOTIN position, $b_z<0$, whereas for shorter tip distances, $b_z>0$. For the chosen value of $J_f$ and the tip located above the NOTIN point, we obtain a positive value of $x_Z$. However, in the absence of the Fe atom, $x_Z$ may in general change sign. This difference becomes important in the limit $J_t/x_Z\rightarrow0$, since the limiting value of $\alpha$ depends on the sign of $x_Z$: $\alpha\rightarrow0$ for $x_Z>0$, whereas $\alpha\rightarrow\pi$ for $x_Z<0$. Physically, this corresponds to an interchange of the energy levels associated with the states $|\uparrow\downarrow\rangle$ and $|\downarrow\uparrow\rangle$.

We first consider the case $x_Z>0$: In the weak-coupling regime, $J_t\ll x_Z$, the angle satisfies $\alpha\simeq0$, and the two Ti spins are effectively decoupled. Conversely, in the strong-coupling regime, $J_t\gg x_Z$, one finds $\alpha\simeq\pi/2$, and the eigenstates correspond to coupled Ti states with total spin $S=1$. In the following, we refer to these two regimes as the \emph{weak-coupling} and \emph{strong-coupling} limits, respectively.



In the rotated basis, the complete Hamiltonian reads: 

\begin{equation}\label{hrot2_reordered_1423}
H_{rot} =
\left(
\begin{array}{cc|cc}
J_{t} + J_f \frac{m_{\rm Fe}}{2} +\sqrt{2}\Delta_b (b_{z} +2B_z) & \sqrt{2}\Delta_b  B_{1,+} & \sqrt{2}\Delta_b B_{1,-} & 0 \\ 
 \sqrt{2}\Delta_b  B_{1,+} & \frac{J_{t}}{2} +\frac{\Gamma}{2}   & 0 & \sqrt{2}\Delta_b B_{-1,+} \\  \hline
 \sqrt{2}\Delta_b B_{1,-} & 0 & \frac{J_{t}}{2} -\frac{\Gamma}{2}  & \sqrt{2}\Delta_b  B_{-1,-} \\ 
 0 & \sqrt{2}\Delta_b B_{-1,+} & \sqrt{2}\Delta_b  B_{-1,-} &  -\sqrt{2}\Delta_b( b_{z} + 2B_z) + J_{t} - J_f \frac{m_{\rm Fe}}{2}
\end{array}
\right)
\end{equation}

\noindent with

\begin{eqnarray}
B_{1,+} = \cos{\left(\frac{\alpha}{2}\right)} B_x + \sin{\left(\frac{\alpha}{2}\right)} (b_x + B_x ) \\
B_{1,-} = \cos{\left(\frac{\alpha}{2}\right)} (b_x +B_x) -  \sin{\left(\frac{\alpha}{2}\right)} B_x  \\ 
B_{-1,+} = \cos{\left(\frac{\alpha}{2}\right)} (b_x +B_x) +  \sin{\left(\frac{\alpha}{2}\right)} B_x  \\ 
B_{-1,-} = \cos{\left(\frac{\alpha}{2}\right)}B_x  -  (b_x +B_x)\sin{\left(\frac{\alpha}{2}\right)}
\end{eqnarray} 

For each projection $S^{(z)}_{\rm Fe}$, the ground state is the triplet state with the lowest spin projection, $T_-$. The first excited state is a linear combination of the two Ti-Ti configurations with total projection $S^{(z)}=0$. In the weak-coupling limit, the dominant component of this state is determined by the sign of $x_Z$: for $x_Z>0$, the configuration $|\downarrow,\uparrow\rangle$ is favored over $|\uparrow,\downarrow\rangle$. Therefore, the control mechanism can be understood as being driven by a current through the first Ti atom, which remains predominantly in the $|\uparrow\rangle$ configuration, while the second Ti spin is flipped.


On the other hand, in the strong-coupling limit, the first excited state is the singlet $S$, and the sign of $x_Z$ is less relevant than in the weak-coupling case. Depending on the ratio $J_t/x_Z$, two different control regimes can be identified. For strong coupling, the dynamics mainly involves transitions between the singlet $S$ and the $T_-$ triplet state. In contrast, in the weak-coupling regime, control can be achieved through a spin flip of one of the Ti atoms, mediated by the Fe atom.

The Hamiltonian for the two-level system formed with the lowest energy states is:  

\begin{equation}
H^{TLS}_{rot} =
\left(
\begin{array}{cc}
  \frac{J_{t}}{2} -\frac{\Gamma}{2}  & \sqrt{2}\Delta_b  B_{-1,-} \\ 
\sqrt{2}\Delta_b  B_{-1,-} &  -\sqrt{2}\Delta_b( b_{z} + 2B_z) + J_{t} - J_f \frac{m_{\rm Fe}}{2}
\end{array}
\right)
\end{equation}

In the experiment, by setting a particular value of the magnetic field $B_z$ with $B_x=0$, we could set a degeneracy condition for these two levels. This allows for creating an avoided crossing for this specific value of $B_z$ with $B_x\neq0$. We assume that $m_{Fe}=-2$, since that projection corresponds to the ground state multiplet of the Fe atom when $B_z>0$. 

The applied electric field induces changes in the system. In particular, we assume that $b_z \simeq b_z^{(0)}+\delta b_z$, where $b_z^{(0)}$ is the tip magnetic-field component in the absence of an applied voltage and $\delta b_z$ is the voltage-induced modulation. Similarly, we write $J_f \simeq J_f^{(0)}+\delta J_f$. These changes are then propagated to the quantities $x_Z$, $\Gamma$, and $\alpha$:

\begin{eqnarray}
\delta x_Z&=& 2\sqrt{2}\mu_B \delta b^{(z)}_0 + 2\delta J_f \\ 
\delta \Gamma &=& \frac{x_Z}{\Gamma}\delta x_Z = \frac{x_Z}{\Gamma} \left( g\mu_B \delta b^{(z)}_0 + 2 \delta J_f \right) \\ 
\delta \alpha &=& -\frac{J_t}{\Gamma^2} \delta x_Z 
\end{eqnarray}

From Eq.~\eqref{htls-g}, setting $B_x=0$, we can estimate the value of $B_z$ required to satisfy the degeneracy condition at $V_{dc}=0$:

\begin{eqnarray}
B_z = -\frac{b^{(0)}_z}{2}+ \frac{J_t +2 J^{(0)}_f  + \Gamma}{4\sqrt{2}\Delta_b} 
\end{eqnarray}

\noindent For $B_x\neq 0$, the two states become coupled by an off-diagonal term of the form $\sqrt{2}\Delta_b B{-1,-}$.


When analyzing LZSM dynamics, the effective two-level Hamiltonian is usually written in the standard form:

\begin{equation}
 H^{TLS}_{rot} = \frac{\varepsilon}{2} \sigma_z +\frac{\Delta}{2} \sigma_x     
\end{equation}

\noindent and, in our particular case, we have: 

\begin{eqnarray}
 \varepsilon &=& -\frac{\left(J_t+\Gamma\right)}{2} +\sqrt{2}\Delta_b b_z + 2\sqrt{2}\Delta_b B_z - J_f  \\
 \Delta &=& 2\sqrt{2} \Delta_b B_{-1,-} 
\end{eqnarray}

The expression above can be applied generally and, in principle, may have a nonlinear dependence on the applied voltage. We now introduce a simplifying assumption: the dependence of both $\varepsilon$ and $\Delta$ on the voltage-induced changes in $J_f$ and $b_z$ can be accurately described by a first-order expansion. This approximation is not rigorously justified in all regimes and may therefore break down in some parameter ranges.


Let us now consider the small-voltage limit and assume that the magnetic field is chosen such that the two levels are degenerate at $V_{\rm dc}=0$, as indicated by Eq.~\eqref{bcross}. Under this condition, evaluating $J_f$ and $b_z$ at $V_{\rm dc}=0$ in Eq.~\eqref{eps2level} gives exactly $\varepsilon^{(0)}=0$. Therefore, the first-order expansion around $V_{\rm dc}=0$ reads:

\begin{equation}
\varepsilon \simeq \frac{\partial \varepsilon }{\partial  b_z} \delta b_z +\frac{\partial \varepsilon }{\partial  J_f} \delta J_f ,      
\end{equation}

\noindent the explicit form is :

\begin{eqnarray}
\varepsilon &=& \left( 1- \frac{ x_Z^{(0)}}{\Gamma^{(0)}} \right) \sqrt{2}\Delta_b \delta b_z  - \left( 1+ \frac{ x_Z^{(0)}}{\Gamma^{(0)}} \right)\delta J_f   
\end{eqnarray}

\noindent On the other hand $\Delta=\Delta^{(0)}+\delta\Delta$ with

\begin{eqnarray}
\Delta^{(0)}&=& \sqrt{2} \Delta_b \left( 2\cos{\left(\frac{\alpha_0}{2}\right)}B_x  -  2(b_x +B_x)\sin{\left(\frac{\alpha_0}{2}\right)} \right)\\
\delta \Delta &=& \sqrt{2} \Delta_b  \left(  \sin{\left(\frac{\alpha_0}{2}\right)}B_x  +  (b_x +B_x)\cos{\left(\frac{\alpha_0}{2}\right)} \right) \frac{J_t}{(\Gamma^{(0)})^2} \delta x_Z 
\end{eqnarray}

The expression above is, in principle, quite general and remains valid for arbitrary values of the coupling and magnetic field, as long as both $B_x$ and $b_x$ can be treated as small perturbations.


In the weak-coupling limit with $x_Z>0$, we can start from the zeroth-order approximation, for which $\tan(\alpha)\simeq\alpha\simeq0$. In this limit, $\Gamma\simeq x_Z$, and therefore $B_{-1,-}\simeq B_x$. Moreover, since $d\alpha/dx_Z$ is proportional to $J_t/\Gamma^2\simeq J_t/x_Z^2\simeq0$, the dependence of $\alpha$ on $x_Z$ can be safely neglected. Thus, we can state that:

\begin{eqnarray}
\varepsilon &\simeq& - 2 \delta J_f \\
\Delta &\simeq& g \mu_B B_x  \\
g \mu_B B_z &=& 2J^{(0)}_f  + \frac{J_t}{2}
\end{eqnarray}

This equation shows that, when the weak-coupling condition is satisfied, namely $J_t\ll x_Z$, the system can be controlled through $J_f$. In other words, the Fe atom provides a way to control the Ti state remotely. In this limit, the dynamics takes place within the subspace ${|\downarrow\uparrow\rangle,|\downarrow\downarrow\rangle}$, and the corresponding gap is determined by the magnetic field.





\subsection{ The Ti-Ti dimmer without the Fe atom }

For $J_f=0$, we have

\[
\tan\alpha=\frac{J_t}{g\mu_B b_z}.
\]

\noindent In this case, it is useful to distinguish between two regimes. For tip positions above the NOTIN point, as in the case of a predominantly dipolar tip, one obtains $x_Z<0$. For an exchange-dominated tip, the contribution changes sign, leading to $x_Z>0$.

\subsubsection{Remote Switching of a Distant Ti Adatom via STM}

Here we consider the weak-coupling regime, $J_t/x_Z \ll 1$. For an exchange tip, for which $x_Z>0$, the ground state is approximately $|\downarrow,\downarrow\rangle$, while the first excited state is approximately $|\downarrow,\uparrow\rangle$. The effective two-level Hamiltonian, written in terms of the detuning $\varepsilon$ and the gap $\Delta$, is given by

\begin{eqnarray} \label{epsjf0}
\varepsilon &=& \left( 1- \frac{ 1}{\sqrt{ 1 +\tan{\alpha}^2 }} \right) \sqrt{2}\Delta_b \delta b_z   \\ 
\Delta &\simeq & \frac{B_x}{\sqrt{2}}\left( \sqrt{1-\frac{1}{\sqrt{1 +\tan{\alpha}^2}}} -\sqrt{1 + \frac{1}{\sqrt{1 +\tan{\alpha}^2}}}  \right)  - \frac{b_x}{\sqrt{2}}\sqrt{1-\frac{1}{\sqrt{1 +\tan{\alpha}^2}}}
\end{eqnarray}

\noindent Near the weak coupling limit, we only keep the leading order terms in $\tan{\alpha}$ :

\begin{eqnarray} \label{epsjf0xz+}
\varepsilon &=& \left( \frac{ J_t}{g \mu_B b^{(0)}_z} \right)^2 \frac{\Delta_b}{\sqrt{2}} \delta b_z   \\ 
\Delta &\simeq & - \sqrt{2}\Delta_b B_x\left( 1- \frac{1}{\sqrt{2}}\left( \frac{ J_t}{g \mu_B b^{(0)}_z} \right)  \right) - \Delta_b b_x\left( \frac{ J_t}{g \mu_B b^{(0)}_z} \right) \label{dpsjf0xz+}
\end{eqnarray}

In this expression, the coupling $\Delta$ is approximately given by the $x$ component of the external magnetic field, $B_x$, with only a very small contribution from $b_x$. In addition, the control amplitude is proportional to $(J_t/x_Z)^2$, which is small within the present approximation. This indicates that, in this regime, tip-induced control in the absence of the nearby Fe atom would be rather difficult, since it would require larger voltage amplitudes.

\subsubsection{Local Switching of a Ti Adatom Beneath the STM Tip}

We continue to consider the weak-coupling regime, $J_t/|x_Z|\ll 1$, but now for the case $x_Z<0$. This situation corresponds to tip distances larger than the NOTIN point. In this regime, the sign change of $x_Z$ leads to an inversion of the original level structure. In particular, instead of having $|\downarrow,\uparrow\rangle$ as the first excited state, the first excited state becomes approximately $|\uparrow,\downarrow\rangle$.

Therefore, the dynamics in this case corresponds essentially to the control of the atom located underneath the tip, as if it were an isolated spin, within the subspace $|\uparrow,\downarrow\rangle \rightarrow |\downarrow,\downarrow\rangle$. Since $x_Z<0$, the rotation angle tends to $\alpha=\pi$, rather than to zero, in the limit $J_t\rightarrow 0$. This behavior is ultimately responsible for the inversion of the levels. We can then write $\alpha=\pi-\beta$, so that the weak-coupling limit corresponds to $\beta\rightarrow 0$. Using $\beta$ as the expansion parameter, a Taylor expansion can be performed. It is straightforward to show that, in this limit,
$\tan\beta = J_t/|x_Z|$, and

\begin{eqnarray}
\varepsilon &=& \left( 1+ \frac{ 1}{\sqrt{ 1 +\tan{\beta}^2 }} \right) \sqrt{2}\Delta_b \delta b^{(0)}_z   \\ 
B_{-1,-} &=& \cos{\left(\frac{\beta}{2}\right)} (b_x +B_x) +  \sin{\left(\frac{\beta}{2}\right)} B_x  \\ 
\end{eqnarray}

\begin{eqnarray} \label{epsjf0xz-}
\varepsilon &=& \left( 1+ \left( \frac{ J_t}{g \mu_B b^{(0)}_z} \right)^2 \right) \frac{\Delta_b}{\sqrt{2}} \delta b^{(0)}_z   \\ 
\Delta &\simeq & - \Delta_b (B_x+b_x) - \Delta_b B_x\left( \frac{ J_t}{g \mu_B b^{(0)}_z} \right) \label{dpsjf0xz-}
\end{eqnarray}

Although coherent control is possible in this case, the second Ti atom does not play an active role in the dynamics. The system therefore behaves essentially as an isolated Ti atom driven by the STM tip. In this regime, however, the anisotropy of the tip plays an important role in opening the gap. As a consequence, the gap is not easy to tune experimentally.








\section{LZSM in the fast driving regime}\label{Alzsm}

The essential physics can be described in terms of a driven two-level system. In the diabatic basis\cite{nori1}, the Hamiltonian can be written as \cite{nori1,nori2}

\begin{equation}
H_{\mathrm{TLS}}(t)=-\frac{1}{2}\left(\varepsilon(t) \sigma_z + \Delta \sigma_x \right),
\label{eq:TLS_LZSM}
\end{equation}

\noindent where $\varepsilon(t)$ is the detuning and $\Delta$ is the minimum gap at the avoided crossing. We consider a harmonic modulation, $\varepsilon(t)=\varepsilon_0 + A \sin(\omega t)$, where $\varepsilon_0$ is the dc detuning, $A$ is the driving amplitude, and $\omega$ is the driving frequency. If the modulation amplitude is large enough for $\varepsilon(t)$ to change sign during each cycle, the system is repeatedly driven across the avoided crossing. Each passage then acts as a quantum beam splitter. The probability for a diabatic transition in a single passage is given by the Landau-Zener formula \cite{nori1}

\begin{equation}
P_{\mathrm{LZ}}= \exp \left( -2\pi \delta \right),
\qquad
\delta = \frac{\Delta^2}{4 v},
\label{eq:PLZ}
\end{equation}

\noindent where $v=|\dot{\varepsilon}(t_c)|$ is the sweep velocity at the crossing time $t_c$, defined by $\varepsilon(t_c)=0$. The phase accumulated between consecutive passages determines whether the interference is constructive or destructive. Therefore, population transfer is controlled not only by the size of the gap $\Delta$, but also by the driving amplitude, frequency, and dc detuning.

Of particular interest is the fast-driving regime, defined by $\delta = \Delta^2/(4v)\ll 1$, where each individual passage through the avoided crossing is strongly nonadiabatic. In this regime, the system remains mostly in the same diabatic state after a single passage. However, repeated passages through the avoided crossing can still lead to population transfer due to quantum interference \cite{nori1,nori2}.

The transition probability between diabatic states can be written as a sum of resonant contributions associated with the different multiphoton conditions $\varepsilon_0 \simeq n\omega$. Close to the $n$-th resonance, the probability takes the form

\begin{eqnarray}\label{fd1}
P^{(n)}_{-\rightarrow +}(t)&=& \frac{\Gamma_n^2}{2\Omega_n^2}\left(1-\cos{(\Omega_n t)}\right)   \nonumber \\
\Omega_n&=&\sqrt{\left(n\omega-\varepsilon_0\right)^2+\Gamma_n^2}
\end{eqnarray}

\noindent where $\Gamma_n=\Delta J_n(A/\omega)$ and $J_n$ are the Bessel functions of the first kind. This expression shows that population transfer can be enhanced or suppressed by changing the ratio $A/\omega$. In particular, when $J_n(A/\omega)=0$, the transition probability is suppressed even in the presence of a finite avoided-crossing gap $\Delta$, corresponding to coherent destruction of tunneling.


\bibliography{biblio}{}

\begin{thebibliography}{34}%
\makeatletter
\providecommand \@ifxundefined [1]{%
 \@ifx{#1\undefined}
}%
\providecommand \@ifnum [1]{%
 \ifnum #1\expandafter \@firstoftwo
 \else \expandafter \@secondoftwo
 \fi
}%
\providecommand \@ifx [1]{%
 \ifx #1\expandafter \@firstoftwo
 \else \expandafter \@secondoftwo
 \fi
}%
\providecommand \natexlab [1]{#1}%
\providecommand \enquote  [1]{``#1''}%
\providecommand \bibnamefont  [1]{#1}%
\providecommand \bibfnamefont [1]{#1}%
\providecommand \citenamefont [1]{#1}%
\providecommand \href@noop [0]{\@secondoftwo}%
\providecommand \href [0]{\begingroup \@sanitize@url \@href}%
\providecommand \@href[1]{\@@startlink{#1}\@@href}%
\providecommand \@@href[1]{\endgroup#1\@@endlink}%
\providecommand \@sanitize@url [0]{\catcode `\\12\catcode `\$12\catcode
  `\&12\catcode `\#12\catcode `\^12\catcode `\_12\catcode `\%12\relax}%
\providecommand \@@startlink[1]{}%
\providecommand \@@endlink[0]{}%
\providecommand \url  [0]{\begingroup\@sanitize@url \@url }%
\providecommand \@url [1]{\endgroup\@href {#1}{\urlprefix }}%
\providecommand \urlprefix  [0]{URL }%
\providecommand \Eprint [0]{\href }%
\providecommand \doibase [0]{https://doi.org/}%
\providecommand \selectlanguage [0]{\@gobble}%
\providecommand \bibinfo  [0]{\@secondoftwo}%
\providecommand \bibfield  [0]{\@secondoftwo}%
\providecommand \translation [1]{[#1]}%
\providecommand \BibitemOpen [0]{}%
\providecommand \bibitemStop [0]{}%
\providecommand \bibitemNoStop [0]{.\EOS\space}%
\providecommand \EOS [0]{\spacefactor3000\relax}%
\providecommand \BibitemShut  [1]{\csname bibitem#1\endcsname}%
\let\auto@bib@innerbib\@empty
\bibitem [{\citenamefont {Baumann}\ \emph {et~al.}(2015)\citenamefont
  {Baumann}, \citenamefont {Paul}, \citenamefont {Choi}, \citenamefont {Lutz},
  \citenamefont {Ardavan},\ and\ \citenamefont {Heinrich}}]{Baumann2015}%
  \BibitemOpen
  \bibfield  {author} {\bibinfo {author} {\bibfnamefont {S.}~\bibnamefont
  {Baumann}}, \bibinfo {author} {\bibfnamefont {W.}~\bibnamefont {Paul}},
  \bibinfo {author} {\bibfnamefont {T.}~\bibnamefont {Choi}}, \bibinfo {author}
  {\bibfnamefont {C.~P.}\ \bibnamefont {Lutz}}, \bibinfo {author}
  {\bibfnamefont {A.}~\bibnamefont {Ardavan}},\ and\ \bibinfo {author}
  {\bibfnamefont {A.~J.}\ \bibnamefont {Heinrich}},\ }\bibfield  {title}
  {\bibinfo {title} {Electron paramagnetic resonance of individual atoms on a
  surface},\ }\href@noop {} {\bibfield  {journal} {\bibinfo  {journal}
  {Science}\ }\textbf {\bibinfo {volume} {350}},\ \bibinfo {pages} {417}
  (\bibinfo {year} {2015})}\BibitemShut {NoStop}%
\bibitem [{\citenamefont {Taherpour}\ \emph {et~al.}(2026)\citenamefont
  {Taherpour}, \citenamefont {Jankovi{\'c}}, \citenamefont {Reina-Galvez},
  \citenamefont {Le},\ and\ \citenamefont {Wolf}}]{taherpour2026review-esr}%
  \BibitemOpen
  \bibfield  {author} {\bibinfo {author} {\bibfnamefont {S.}~\bibnamefont
  {Taherpour}}, \bibinfo {author} {\bibfnamefont {D.}~\bibnamefont
  {Jankovi{\'c}}}, \bibinfo {author} {\bibfnamefont {J.}~\bibnamefont
  {Reina-Galvez}}, \bibinfo {author} {\bibfnamefont {H.-A.}\ \bibnamefont
  {Le}},\ and\ \bibinfo {author} {\bibfnamefont {C.}~\bibnamefont {Wolf}},\
  }\bibfield  {title} {\bibinfo {title} {Theory of electron spin resonance
  scanning tunneling microscopy: the first decade},\ }\href@noop {} {\bibfield
  {journal} {\bibinfo  {journal} {Journal of Physics: Condensed Matter}\
  }\textbf {\bibinfo {volume} {38}},\ \bibinfo {pages} {243003} (\bibinfo
  {year} {2026})}\BibitemShut {NoStop}%
\bibitem [{\citenamefont {Willke}\ \emph {et~al.}(2021)\citenamefont {Willke},
  \citenamefont {Bilgeri}, \citenamefont {Zhang}, \citenamefont {Wang},
  \citenamefont {Wolf}, \citenamefont {Aubin}, \citenamefont {Heinrich},\ and\
  \citenamefont {Choi}}]{willke2021a}%
  \BibitemOpen
  \bibfield  {author} {\bibinfo {author} {\bibfnamefont {P.}~\bibnamefont
  {Willke}}, \bibinfo {author} {\bibfnamefont {T.}~\bibnamefont {Bilgeri}},
  \bibinfo {author} {\bibfnamefont {X.}~\bibnamefont {Zhang}}, \bibinfo
  {author} {\bibfnamefont {Y.}~\bibnamefont {Wang}}, \bibinfo {author}
  {\bibfnamefont {C.}~\bibnamefont {Wolf}}, \bibinfo {author} {\bibfnamefont
  {H.}~\bibnamefont {Aubin}}, \bibinfo {author} {\bibfnamefont
  {A.}~\bibnamefont {Heinrich}},\ and\ \bibinfo {author} {\bibfnamefont
  {T.}~\bibnamefont {Choi}},\ }\bibfield  {title} {\bibinfo {title} {Coherent
  spin control of single molecules on a surface},\ }\href@noop {} {\bibfield
  {journal} {\bibinfo  {journal} {ACS nano}\ }\textbf {\bibinfo {volume}
  {15}},\ \bibinfo {pages} {17959} (\bibinfo {year} {2021})}\BibitemShut
  {NoStop}%
\bibitem [{\citenamefont {Wang}\ \emph
  {et~al.}(2023{\natexlab{a}})\citenamefont {Wang}, \citenamefont {Chen},
  \citenamefont {Bui}, \citenamefont {Wolf}, \citenamefont {Haze},
  \citenamefont {Mier}, \citenamefont {Kim}, \citenamefont {Choi},
  \citenamefont {Lutz}, \citenamefont {Bae} \emph {et~al.}}]{wang2023qb}%
  \BibitemOpen
  \bibfield  {author} {\bibinfo {author} {\bibfnamefont {Y.}~\bibnamefont
  {Wang}}, \bibinfo {author} {\bibfnamefont {Y.}~\bibnamefont {Chen}}, \bibinfo
  {author} {\bibfnamefont {H.~T.}\ \bibnamefont {Bui}}, \bibinfo {author}
  {\bibfnamefont {C.}~\bibnamefont {Wolf}}, \bibinfo {author} {\bibfnamefont
  {M.}~\bibnamefont {Haze}}, \bibinfo {author} {\bibfnamefont {C.}~\bibnamefont
  {Mier}}, \bibinfo {author} {\bibfnamefont {J.}~\bibnamefont {Kim}}, \bibinfo
  {author} {\bibfnamefont {D.-J.}\ \bibnamefont {Choi}}, \bibinfo {author}
  {\bibfnamefont {C.~P.}\ \bibnamefont {Lutz}}, \bibinfo {author}
  {\bibfnamefont {Y.}~\bibnamefont {Bae}}, \emph {et~al.},\ }\bibfield  {title}
  {\bibinfo {title} {An atomic-scale multi-qubit platform},\ }\href@noop {}
  {\bibfield  {journal} {\bibinfo  {journal} {Science}\ }\textbf {\bibinfo
  {volume} {382}},\ \bibinfo {pages} {87} (\bibinfo {year}
  {2023}{\natexlab{a}})}\BibitemShut {NoStop}%
\bibitem [{\citenamefont {Willke}\ \emph
  {et~al.}(2018{\natexlab{a}})\citenamefont {Willke}, \citenamefont {Paul},
  \citenamefont {Natterer}, \citenamefont {Yang}, \citenamefont {Bae},
  \citenamefont {Choi}, \citenamefont {Fern{\'a}ndez-Rossier}, \citenamefont
  {Heinrich},\ and\ \citenamefont {Lutz}}]{willke2018}%
  \BibitemOpen
  \bibfield  {author} {\bibinfo {author} {\bibfnamefont {P.}~\bibnamefont
  {Willke}}, \bibinfo {author} {\bibfnamefont {W.}~\bibnamefont {Paul}},
  \bibinfo {author} {\bibfnamefont {F.~D.}\ \bibnamefont {Natterer}}, \bibinfo
  {author} {\bibfnamefont {K.}~\bibnamefont {Yang}}, \bibinfo {author}
  {\bibfnamefont {Y.}~\bibnamefont {Bae}}, \bibinfo {author} {\bibfnamefont
  {T.}~\bibnamefont {Choi}}, \bibinfo {author} {\bibfnamefont {J.}~\bibnamefont
  {Fern{\'a}ndez-Rossier}}, \bibinfo {author} {\bibfnamefont {A.~J.}\
  \bibnamefont {Heinrich}},\ and\ \bibinfo {author} {\bibfnamefont {C.~P.}\
  \bibnamefont {Lutz}},\ }\bibfield  {title} {\bibinfo {title} {Probing quantum
  coherence in single-atom electron spin resonance},\ }\href@noop {} {\bibfield
   {journal} {\bibinfo  {journal} {Science advances}\ }\textbf {\bibinfo
  {volume} {4}},\ \bibinfo {pages} {eaaq1543} (\bibinfo {year}
  {2018}{\natexlab{a}})}\BibitemShut {NoStop}%
\bibitem [{\citenamefont {Willke}\ \emph
  {et~al.}(2018{\natexlab{b}})\citenamefont {Willke}, \citenamefont {Bae},
  \citenamefont {Yang}, \citenamefont {Lado}, \citenamefont {Ferr{\'o}n},
  \citenamefont {Choi}, \citenamefont {Ardavan}, \citenamefont
  {Fern{\'a}ndez-Rossier}, \citenamefont {Heinrich},\ and\ \citenamefont
  {Lutz}}]{willke2018b}%
  \BibitemOpen
  \bibfield  {author} {\bibinfo {author} {\bibfnamefont {P.}~\bibnamefont
  {Willke}}, \bibinfo {author} {\bibfnamefont {Y.}~\bibnamefont {Bae}},
  \bibinfo {author} {\bibfnamefont {K.}~\bibnamefont {Yang}}, \bibinfo {author}
  {\bibfnamefont {J.~L.}\ \bibnamefont {Lado}}, \bibinfo {author}
  {\bibfnamefont {A.}~\bibnamefont {Ferr{\'o}n}}, \bibinfo {author}
  {\bibfnamefont {T.}~\bibnamefont {Choi}}, \bibinfo {author} {\bibfnamefont
  {A.}~\bibnamefont {Ardavan}}, \bibinfo {author} {\bibfnamefont
  {J.}~\bibnamefont {Fern{\'a}ndez-Rossier}}, \bibinfo {author} {\bibfnamefont
  {A.~J.}\ \bibnamefont {Heinrich}},\ and\ \bibinfo {author} {\bibfnamefont
  {C.~P.}\ \bibnamefont {Lutz}},\ }\bibfield  {title} {\bibinfo {title}
  {Hyperfine interaction of individual atoms on a surface},\ }\href@noop {}
  {\bibfield  {journal} {\bibinfo  {journal} {Science}\ }\textbf {\bibinfo
  {volume} {362}},\ \bibinfo {pages} {336} (\bibinfo {year}
  {2018}{\natexlab{b}})}\BibitemShut {NoStop}%
\bibitem [{\citenamefont {Willke}\ \emph
  {et~al.}(2019{\natexlab{a}})\citenamefont {Willke}, \citenamefont {Yang},
  \citenamefont {Bae}, \citenamefont {Heinrich},\ and\ \citenamefont
  {Lutz}}]{willke2019}%
  \BibitemOpen
  \bibfield  {author} {\bibinfo {author} {\bibfnamefont {P.}~\bibnamefont
  {Willke}}, \bibinfo {author} {\bibfnamefont {K.}~\bibnamefont {Yang}},
  \bibinfo {author} {\bibfnamefont {Y.}~\bibnamefont {Bae}}, \bibinfo {author}
  {\bibfnamefont {A.~J.}\ \bibnamefont {Heinrich}},\ and\ \bibinfo {author}
  {\bibfnamefont {C.~P.}\ \bibnamefont {Lutz}},\ }\bibfield  {title} {\bibinfo
  {title} {Magnetic resonance imaging of single atoms on a surface},\
  }\href@noop {} {\bibfield  {journal} {\bibinfo  {journal} {Nature Physics}\
  ,\ \bibinfo {pages} {1}} (\bibinfo {year} {2019}{\natexlab{a}})}\BibitemShut
  {NoStop}%
\bibitem [{\citenamefont {Willke}\ \emph
  {et~al.}(2019{\natexlab{b}})\citenamefont {Willke}, \citenamefont {Singha},
  \citenamefont {Zhang}, \citenamefont {Esat}, \citenamefont {Lutz},
  \citenamefont {Heinrich},\ and\ \citenamefont {Choi}}]{willke2019b}%
  \BibitemOpen
  \bibfield  {author} {\bibinfo {author} {\bibfnamefont {P.}~\bibnamefont
  {Willke}}, \bibinfo {author} {\bibfnamefont {A.}~\bibnamefont {Singha}},
  \bibinfo {author} {\bibfnamefont {X.}~\bibnamefont {Zhang}}, \bibinfo
  {author} {\bibfnamefont {T.}~\bibnamefont {Esat}}, \bibinfo {author}
  {\bibfnamefont {C.~P.}\ \bibnamefont {Lutz}}, \bibinfo {author}
  {\bibfnamefont {A.~J.}\ \bibnamefont {Heinrich}},\ and\ \bibinfo {author}
  {\bibfnamefont {T.}~\bibnamefont {Choi}},\ }\bibfield  {title} {\bibinfo
  {title} {Tuning single-atom electron spin resonance in a vector-magnetic
  field},\ }\href@noop {} {\bibfield  {journal} {\bibinfo  {journal} {arXiv
  preprint arXiv:1908.11061}\ } (\bibinfo {year}
  {2019}{\natexlab{b}})}\BibitemShut {NoStop}%
\bibitem [{\citenamefont {Yang}\ \emph {et~al.}(2017)\citenamefont {Yang},
  \citenamefont {Bae}, \citenamefont {Paul}, \citenamefont {Natterer},
  \citenamefont {Willke}, \citenamefont {Lado}, \citenamefont {Ferr{\'o}n},
  \citenamefont {Choi}, \citenamefont {Fern{\'a}ndez-Rossier}, \citenamefont
  {Heinrich} \emph {et~al.}}]{yang2017}%
  \BibitemOpen
  \bibfield  {author} {\bibinfo {author} {\bibfnamefont {K.}~\bibnamefont
  {Yang}}, \bibinfo {author} {\bibfnamefont {Y.}~\bibnamefont {Bae}}, \bibinfo
  {author} {\bibfnamefont {W.}~\bibnamefont {Paul}}, \bibinfo {author}
  {\bibfnamefont {F.~D.}\ \bibnamefont {Natterer}}, \bibinfo {author}
  {\bibfnamefont {P.}~\bibnamefont {Willke}}, \bibinfo {author} {\bibfnamefont
  {J.~L.}\ \bibnamefont {Lado}}, \bibinfo {author} {\bibfnamefont
  {A.}~\bibnamefont {Ferr{\'o}n}}, \bibinfo {author} {\bibfnamefont
  {T.}~\bibnamefont {Choi}}, \bibinfo {author} {\bibfnamefont {J.}~\bibnamefont
  {Fern{\'a}ndez-Rossier}}, \bibinfo {author} {\bibfnamefont {A.~J.}\
  \bibnamefont {Heinrich}}, \emph {et~al.},\ }\bibfield  {title} {\bibinfo
  {title} {Engineering the eigenstates of coupled spin-1/2 atoms on a
  surface},\ }\href@noop {} {\bibfield  {journal} {\bibinfo  {journal}
  {Physical review letters}\ }\textbf {\bibinfo {volume} {119}},\ \bibinfo
  {pages} {227206} (\bibinfo {year} {2017})}\BibitemShut {NoStop}%
\bibitem [{\citenamefont {Seifert}\ \emph {et~al.}(2019)\citenamefont
  {Seifert}, \citenamefont {Kovarik}, \citenamefont {Nistor}, \citenamefont
  {Persichetti}, \citenamefont {Stepanow},\ and\ \citenamefont
  {Gambardella}}]{seifert2019}%
  \BibitemOpen
  \bibfield  {author} {\bibinfo {author} {\bibfnamefont {T.}~\bibnamefont
  {Seifert}}, \bibinfo {author} {\bibfnamefont {S.}~\bibnamefont {Kovarik}},
  \bibinfo {author} {\bibfnamefont {C.}~\bibnamefont {Nistor}}, \bibinfo
  {author} {\bibfnamefont {L.}~\bibnamefont {Persichetti}}, \bibinfo {author}
  {\bibfnamefont {S.}~\bibnamefont {Stepanow}},\ and\ \bibinfo {author}
  {\bibfnamefont {P.}~\bibnamefont {Gambardella}},\ }\bibfield  {title}
  {\bibinfo {title} {Single-atom electron paramagnetic resonance in a scanning
  tunneling microscope driven by a radiofrequency antenna at 4 k},\ }\href@noop
  {} {\bibfield  {journal} {\bibinfo  {journal} {arXiv preprint
  arXiv:1908.03379}\ } (\bibinfo {year} {2019})}\BibitemShut {NoStop}%
\bibitem [{\citenamefont {Seifert}\ \emph {et~al.}(2021)\citenamefont
  {Seifert}, \citenamefont {Kovarik}, \citenamefont {Gambardella},\ and\
  \citenamefont {Stepanow}}]{seifert2021}%
  \BibitemOpen
  \bibfield  {author} {\bibinfo {author} {\bibfnamefont {T.~S.}\ \bibnamefont
  {Seifert}}, \bibinfo {author} {\bibfnamefont {S.}~\bibnamefont {Kovarik}},
  \bibinfo {author} {\bibfnamefont {P.}~\bibnamefont {Gambardella}},\ and\
  \bibinfo {author} {\bibfnamefont {S.}~\bibnamefont {Stepanow}},\ }\bibfield
  {title} {\bibinfo {title} {Accurate measurement of atomic magnetic moments by
  minimizing the tip magnetic field in stm-based electron paramagnetic
  resonance},\ }\href@noop {} {\bibfield  {journal} {\bibinfo  {journal}
  {Physical Review Research}\ }\textbf {\bibinfo {volume} {3}},\ \bibinfo
  {pages} {043185} (\bibinfo {year} {2021})}\BibitemShut {NoStop}%
\bibitem [{\citenamefont {Seifert}\ \emph {et~al.}(2020)\citenamefont
  {Seifert}, \citenamefont {Kovarik}, \citenamefont {Juraschek}, \citenamefont
  {Spaldin}, \citenamefont {Gambardella},\ and\ \citenamefont
  {Stepanow}}]{seifert2020}%
  \BibitemOpen
  \bibfield  {author} {\bibinfo {author} {\bibfnamefont {T.~S.}\ \bibnamefont
  {Seifert}}, \bibinfo {author} {\bibfnamefont {S.}~\bibnamefont {Kovarik}},
  \bibinfo {author} {\bibfnamefont {D.~M.}\ \bibnamefont {Juraschek}}, \bibinfo
  {author} {\bibfnamefont {N.~A.}\ \bibnamefont {Spaldin}}, \bibinfo {author}
  {\bibfnamefont {P.}~\bibnamefont {Gambardella}},\ and\ \bibinfo {author}
  {\bibfnamefont {S.}~\bibnamefont {Stepanow}},\ }\bibfield  {title} {\bibinfo
  {title} {Longitudinal and transverse electron paramagnetic resonance in a
  scanning tunneling microscope},\ }\href@noop {} {\bibfield  {journal}
  {\bibinfo  {journal} {Science advances}\ }\textbf {\bibinfo {volume} {6}},\
  \bibinfo {pages} {eabc5511} (\bibinfo {year} {2020})}\BibitemShut {NoStop}%
\bibitem [{\citenamefont {Yang}\ \emph {et~al.}(2019)\citenamefont {Yang},
  \citenamefont {Paul}, \citenamefont {Natterer}, \citenamefont {Lado},
  \citenamefont {Bae}, \citenamefont {Willke}, \citenamefont {Choi},
  \citenamefont {Ferr{\'o}n}, \citenamefont {Fern{\'a}ndez-Rossier},
  \citenamefont {Heinrich} \emph {et~al.}}]{yang2019}%
  \BibitemOpen
  \bibfield  {author} {\bibinfo {author} {\bibfnamefont {K.}~\bibnamefont
  {Yang}}, \bibinfo {author} {\bibfnamefont {W.}~\bibnamefont {Paul}}, \bibinfo
  {author} {\bibfnamefont {F.~D.}\ \bibnamefont {Natterer}}, \bibinfo {author}
  {\bibfnamefont {J.~L.}\ \bibnamefont {Lado}}, \bibinfo {author}
  {\bibfnamefont {Y.}~\bibnamefont {Bae}}, \bibinfo {author} {\bibfnamefont
  {P.}~\bibnamefont {Willke}}, \bibinfo {author} {\bibfnamefont
  {T.}~\bibnamefont {Choi}}, \bibinfo {author} {\bibfnamefont {A.}~\bibnamefont
  {Ferr{\'o}n}}, \bibinfo {author} {\bibfnamefont {J.}~\bibnamefont
  {Fern{\'a}ndez-Rossier}}, \bibinfo {author} {\bibfnamefont {A.~J.}\
  \bibnamefont {Heinrich}}, \emph {et~al.},\ }\bibfield  {title} {\bibinfo
  {title} {Tuning the exchange bias on a single atom from 1 mt to 10 t},\
  }\href@noop {} {\bibfield  {journal} {\bibinfo  {journal} {Physical Review
  Letters}\ }\textbf {\bibinfo {volume} {122}},\ \bibinfo {pages} {227203}
  (\bibinfo {year} {2019})}\BibitemShut {NoStop}%
\bibitem [{\citenamefont {Chen}\ \emph {et~al.}(2023)\citenamefont {Chen},
  \citenamefont {Bae},\ and\ \citenamefont {Heinrich}}]{chen2023harnessing}%
  \BibitemOpen
  \bibfield  {author} {\bibinfo {author} {\bibfnamefont {Y.}~\bibnamefont
  {Chen}}, \bibinfo {author} {\bibfnamefont {Y.}~\bibnamefont {Bae}},\ and\
  \bibinfo {author} {\bibfnamefont {A.~J.}\ \bibnamefont {Heinrich}},\
  }\bibfield  {title} {\bibinfo {title} {Harnessing the quantum behavior of
  spins on surfaces},\ }\href@noop {} {\bibfield  {journal} {\bibinfo
  {journal} {Advanced Materials}\ }\textbf {\bibinfo {volume} {35}},\ \bibinfo
  {pages} {2107534} (\bibinfo {year} {2023})}\BibitemShut {NoStop}%
\bibitem [{\citenamefont {Ivakhnenko}\ \emph {et~al.}(2023)\citenamefont
  {Ivakhnenko}, \citenamefont {Shevchenko},\ and\ \citenamefont
  {Nori}}]{nori1}%
  \BibitemOpen
  \bibfield  {author} {\bibinfo {author} {\bibfnamefont {V.}~\bibnamefont
  {Ivakhnenko}}, \bibinfo {author} {\bibfnamefont {S.~N.}\ \bibnamefont
  {Shevchenko}},\ and\ \bibinfo {author} {\bibfnamefont {F.}~\bibnamefont
  {Nori}},\ }\bibfield  {title} {\bibinfo {title} {Nonadiabatic
  landau--zener--st{\"u}ckelberg--majorana transitions, dynamics, and
  interference},\ }\href@noop {} {\bibfield  {journal} {\bibinfo  {journal}
  {Physics Reports}\ }\textbf {\bibinfo {volume} {995}},\ \bibinfo {pages} {1}
  (\bibinfo {year} {2023})}\BibitemShut {NoStop}%
\bibitem [{\citenamefont {Shevchenko}\ \emph {et~al.}(2010)\citenamefont
  {Shevchenko}, \citenamefont {Ashhab},\ and\ \citenamefont {Nori}}]{nori2}%
  \BibitemOpen
  \bibfield  {author} {\bibinfo {author} {\bibfnamefont {S.~N.}\ \bibnamefont
  {Shevchenko}}, \bibinfo {author} {\bibfnamefont {S.}~\bibnamefont {Ashhab}},\
  and\ \bibinfo {author} {\bibfnamefont {F.}~\bibnamefont {Nori}},\ }\bibfield
  {title} {\bibinfo {title} {Landau--zener--st{\"u}ckelberg interferometry},\
  }\href@noop {} {\bibfield  {journal} {\bibinfo  {journal} {Physics Reports}\
  }\textbf {\bibinfo {volume} {492}},\ \bibinfo {pages} {1} (\bibinfo {year}
  {2010})}\BibitemShut {NoStop}%
\bibitem [{\citenamefont {Campbell}\ \emph {et~al.}(2020)\citenamefont
  {Campbell}, \citenamefont {Shim}, \citenamefont {Kannan}, \citenamefont
  {Winik}, \citenamefont {Kim}, \citenamefont {Melville}, \citenamefont
  {Niedzielski}, \citenamefont {Yoder}, \citenamefont {Tahan}, \citenamefont
  {Gustavsson},\ and\ \citenamefont {Oliver}}]{campbell_2020}%
  \BibitemOpen
  \bibfield  {author} {\bibinfo {author} {\bibfnamefont {D.~L.}\ \bibnamefont
  {Campbell}}, \bibinfo {author} {\bibfnamefont {Y.-P.}\ \bibnamefont {Shim}},
  \bibinfo {author} {\bibfnamefont {B.}~\bibnamefont {Kannan}}, \bibinfo
  {author} {\bibfnamefont {R.}~\bibnamefont {Winik}}, \bibinfo {author}
  {\bibfnamefont {D.~K.}\ \bibnamefont {Kim}}, \bibinfo {author} {\bibfnamefont
  {A.}~\bibnamefont {Melville}}, \bibinfo {author} {\bibfnamefont {B.~M.}\
  \bibnamefont {Niedzielski}}, \bibinfo {author} {\bibfnamefont {J.~L.}\
  \bibnamefont {Yoder}}, \bibinfo {author} {\bibfnamefont {C.}~\bibnamefont
  {Tahan}}, \bibinfo {author} {\bibfnamefont {S.}~\bibnamefont {Gustavsson}},\
  and\ \bibinfo {author} {\bibfnamefont {W.~D.}\ \bibnamefont {Oliver}},\
  }\bibfield  {title} {\bibinfo {title} {Universal nonadiabatic control of
  small-gap superconducting qubits},\ }\href
  {https://doi.org/10.1103/PhysRevX.10.041051} {\bibfield  {journal} {\bibinfo
  {journal} {Phys. Rev. X}\ }\textbf {\bibinfo {volume} {10}},\ \bibinfo
  {pages} {041051} (\bibinfo {year} {2020})}\BibitemShut {NoStop}%
\bibitem [{\citenamefont {C\'aceres}\ \emph {et~al.}(2023)\citenamefont
  {C\'aceres}, \citenamefont {Dom\'{\i}nguez},\ and\ \citenamefont
  {S\'anchez}}]{caceres_2023}%
  \BibitemOpen
  \bibfield  {author} {\bibinfo {author} {\bibfnamefont {J.~J.}\ \bibnamefont
  {C\'aceres}}, \bibinfo {author} {\bibfnamefont {D.}~\bibnamefont
  {Dom\'{\i}nguez}},\ and\ \bibinfo {author} {\bibfnamefont {M.~J.}\
  \bibnamefont {S\'anchez}},\ }\bibfield  {title} {\bibinfo {title} {Fast
  quantum gates based on landau-zener-st\"uckelberg-majorana transitions},\
  }\href {https://doi.org/10.1103/PhysRevA.108.052619} {\bibfield  {journal}
  {\bibinfo  {journal} {Phys. Rev. A}\ }\textbf {\bibinfo {volume} {108}},\
  \bibinfo {pages} {052619} (\bibinfo {year} {2023})}\BibitemShut {NoStop}%
\bibitem [{\citenamefont {Ryzhov}\ \emph {et~al.}(2024)\citenamefont {Ryzhov},
  \citenamefont {Ivakhnenko}, \citenamefont {Shevchenko}, \citenamefont
  {Gonzalez-Zalba},\ and\ \citenamefont {Nori}}]{ryzhov2024}%
  \BibitemOpen
  \bibfield  {author} {\bibinfo {author} {\bibfnamefont {A.~I.}\ \bibnamefont
  {Ryzhov}}, \bibinfo {author} {\bibfnamefont {O.~V.}\ \bibnamefont
  {Ivakhnenko}}, \bibinfo {author} {\bibfnamefont {S.~N.}\ \bibnamefont
  {Shevchenko}}, \bibinfo {author} {\bibfnamefont {M.~F.}\ \bibnamefont
  {Gonzalez-Zalba}},\ and\ \bibinfo {author} {\bibfnamefont {F.}~\bibnamefont
  {Nori}},\ }\bibfield  {title} {\bibinfo {title} {Alternative fast quantum
  logic gates using nonadiabatic landau-zener-st\"uckelberg-majorana
  transitions},\ }\href {https://doi.org/10.1103/PhysRevResearch.6.033340}
  {\bibfield  {journal} {\bibinfo  {journal} {Phys. Rev. Res.}\ }\textbf
  {\bibinfo {volume} {6}},\ \bibinfo {pages} {033340} (\bibinfo {year}
  {2024})}\BibitemShut {NoStop}%
\bibitem [{\citenamefont {Ferreyra}\ \emph {et~al.}(2026)\citenamefont
  {Ferreyra}, \citenamefont {Reparaz}, \citenamefont {Sánchez}, \citenamefont
  {Tosi},\ and\ \citenamefont {Dominguez}}]{ferreyra2026optimizing}%
  \BibitemOpen
  \bibfield  {author} {\bibinfo {author} {\bibfnamefont {S.}~\bibnamefont
  {Ferreyra}}, \bibinfo {author} {\bibfnamefont {V.}~\bibnamefont {Reparaz}},
  \bibinfo {author} {\bibfnamefont {M.~J.}\ \bibnamefont {Sánchez}}, \bibinfo
  {author} {\bibfnamefont {L.}~\bibnamefont {Tosi}},\ and\ \bibinfo {author}
  {\bibfnamefont {D.}~\bibnamefont {Dominguez}},\ }\bibfield  {title} {\bibinfo
  {title} {Optimizing lzsm protocol for high-fidelity gates in open-system
  fluxonium},\ }\href@noop {} {\bibfield  {journal} {\bibinfo  {journal} {arXiv
  preprint arXiv:2607.07999}\ } (\bibinfo {year} {2026})}\BibitemShut {NoStop}%
\bibitem [{\citenamefont {Rodr{\'\i}guez}\ \emph {et~al.}(2025)\citenamefont
  {Rodr{\'\i}guez}, \citenamefont {G{\'o}mez}, \citenamefont
  {Fern{\'a}ndez-Rossier},\ and\ \citenamefont
  {Ferr{\'o}n}}]{rodriguez2024non}%
  \BibitemOpen
  \bibfield  {author} {\bibinfo {author} {\bibfnamefont {S.~A.}\ \bibnamefont
  {Rodr{\'\i}guez}}, \bibinfo {author} {\bibfnamefont {S.~S.}\ \bibnamefont
  {G{\'o}mez}}, \bibinfo {author} {\bibfnamefont {J.}~\bibnamefont
  {Fern{\'a}ndez-Rossier}},\ and\ \bibinfo {author} {\bibfnamefont
  {A.}~\bibnamefont {Ferr{\'o}n}},\ }\bibfield  {title} {\bibinfo {title}
  {Nonresonant electric quantum control of individual on-surface spins},\
  }\href@noop {} {\bibfield  {journal} {\bibinfo  {journal} {Physical Review
  Letters}\ }\textbf {\bibinfo {volume} {134}},\ \bibinfo {pages} {056703}
  (\bibinfo {year} {2025})}\BibitemShut {NoStop}%
\bibitem [{\citenamefont {Wang}\ \emph {et~al.}(2025)\citenamefont {Wang},
  \citenamefont {Chen}, \citenamefont {Fan}, \citenamefont {del Castillo},
  \citenamefont {Ferr{\'o}n}, \citenamefont {Jiang}, \citenamefont {Wu},
  \citenamefont {Li}, \citenamefont {Gao}, \citenamefont {Fan} \emph
  {et~al.}}]{wang2025electrically}%
  \BibitemOpen
  \bibfield  {author} {\bibinfo {author} {\bibfnamefont {H.}~\bibnamefont
  {Wang}}, \bibinfo {author} {\bibfnamefont {J.}~\bibnamefont {Chen}}, \bibinfo
  {author} {\bibfnamefont {P.}~\bibnamefont {Fan}}, \bibinfo {author}
  {\bibfnamefont {Y.}~\bibnamefont {del Castillo}}, \bibinfo {author}
  {\bibfnamefont {A.}~\bibnamefont {Ferr{\'o}n}}, \bibinfo {author}
  {\bibfnamefont {L.}~\bibnamefont {Jiang}}, \bibinfo {author} {\bibfnamefont
  {Z.}~\bibnamefont {Wu}}, \bibinfo {author} {\bibfnamefont {S.}~\bibnamefont
  {Li}}, \bibinfo {author} {\bibfnamefont {H.-J.}\ \bibnamefont {Gao}},
  \bibinfo {author} {\bibfnamefont {H.}~\bibnamefont {Fan}}, \emph {et~al.},\
  }\bibfield  {title} {\bibinfo {title} {Electrically tunable quantum
  interference of atomic spins on surfaces},\ }\href@noop {} {\bibfield
  {journal} {\bibinfo  {journal} {arXiv preprint arXiv:2506.01033}\ } (\bibinfo
  {year} {2025})}\BibitemShut {NoStop}%
\bibitem [{\citenamefont {Bae}\ \emph {et~al.}(2018)\citenamefont {Bae},
  \citenamefont {Yang}, \citenamefont {Willke}, \citenamefont {Choi},
  \citenamefont {Heinrich},\ and\ \citenamefont {Lutz}}]{bae2018}%
  \BibitemOpen
  \bibfield  {author} {\bibinfo {author} {\bibfnamefont {Y.}~\bibnamefont
  {Bae}}, \bibinfo {author} {\bibfnamefont {K.}~\bibnamefont {Yang}}, \bibinfo
  {author} {\bibfnamefont {P.}~\bibnamefont {Willke}}, \bibinfo {author}
  {\bibfnamefont {T.}~\bibnamefont {Choi}}, \bibinfo {author} {\bibfnamefont
  {A.~J.}\ \bibnamefont {Heinrich}},\ and\ \bibinfo {author} {\bibfnamefont
  {C.~P.}\ \bibnamefont {Lutz}},\ }\bibfield  {title} {\bibinfo {title}
  {Enhanced quantum coherence in exchange coupled spins via singlet-triplet
  transitions},\ }\href@noop {} {\bibfield  {journal} {\bibinfo  {journal}
  {Science advances}\ }\textbf {\bibinfo {volume} {4}},\ \bibinfo {pages}
  {eaau4159} (\bibinfo {year} {2018})}\BibitemShut {NoStop}%
\bibitem [{\citenamefont {Kot}\ \emph {et~al.}(2023)\citenamefont {Kot},
  \citenamefont {Ismail}, \citenamefont {Drost}, \citenamefont {Siebrecht},
  \citenamefont {Huang},\ and\ \citenamefont {Ast}}]{kot2023electric}%
  \BibitemOpen
  \bibfield  {author} {\bibinfo {author} {\bibfnamefont {P.}~\bibnamefont
  {Kot}}, \bibinfo {author} {\bibfnamefont {M.}~\bibnamefont {Ismail}},
  \bibinfo {author} {\bibfnamefont {R.}~\bibnamefont {Drost}}, \bibinfo
  {author} {\bibfnamefont {J.}~\bibnamefont {Siebrecht}}, \bibinfo {author}
  {\bibfnamefont {H.}~\bibnamefont {Huang}},\ and\ \bibinfo {author}
  {\bibfnamefont {C.~R.}\ \bibnamefont {Ast}},\ }\bibfield  {title} {\bibinfo
  {title} {Electric control of spin transitions at the atomic scale},\
  }\href@noop {} {\bibfield  {journal} {\bibinfo  {journal} {Nature
  Communications}\ }\textbf {\bibinfo {volume} {14}},\ \bibinfo {pages} {6612}
  (\bibinfo {year} {2023})}\BibitemShut {NoStop}%
\bibitem [{\citenamefont {Phark}\ \emph
  {et~al.}(2023{\natexlab{a}})\citenamefont {Phark}, \citenamefont {Bui},
  \citenamefont {Ferr{\'o}n}, \citenamefont {Fern{\'a}ndez-Rossier},
  \citenamefont {Reina-G{\'a}lvez}, \citenamefont {Wolf}, \citenamefont {Wang},
  \citenamefont {Yang}, \citenamefont {Heinrich},\ and\ \citenamefont
  {Lutz}}]{soohion2023b}%
  \BibitemOpen
  \bibfield  {author} {\bibinfo {author} {\bibfnamefont {S.-h.}\ \bibnamefont
  {Phark}}, \bibinfo {author} {\bibfnamefont {H.~T.}\ \bibnamefont {Bui}},
  \bibinfo {author} {\bibfnamefont {A.}~\bibnamefont {Ferr{\'o}n}}, \bibinfo
  {author} {\bibfnamefont {J.}~\bibnamefont {Fern{\'a}ndez-Rossier}}, \bibinfo
  {author} {\bibfnamefont {J.}~\bibnamefont {Reina-G{\'a}lvez}}, \bibinfo
  {author} {\bibfnamefont {C.}~\bibnamefont {Wolf}}, \bibinfo {author}
  {\bibfnamefont {Y.}~\bibnamefont {Wang}}, \bibinfo {author} {\bibfnamefont
  {K.}~\bibnamefont {Yang}}, \bibinfo {author} {\bibfnamefont {A.~J.}\
  \bibnamefont {Heinrich}},\ and\ \bibinfo {author} {\bibfnamefont {C.~P.}\
  \bibnamefont {Lutz}},\ }\bibfield  {title} {\bibinfo {title}
  {Electric-field-driven spin resonance by on-surface exchange coupling to a
  single-atom magnet},\ }\href@noop {} {\bibfield  {journal} {\bibinfo
  {journal} {Advanced Science}\ }\textbf {\bibinfo {volume} {10}},\ \bibinfo
  {pages} {2302033} (\bibinfo {year} {2023}{\natexlab{a}})}\BibitemShut
  {NoStop}%
\bibitem [{\citenamefont {del Castillo}\ \emph {et~al.}(2026)\citenamefont {del
  Castillo}, \citenamefont {Ferrón},\ and\ \citenamefont
  {Fernández-Rossier}}]{YelkoSP2025}%
  \BibitemOpen
  \bibfield  {author} {\bibinfo {author} {\bibfnamefont {Y.}~\bibnamefont {del
  Castillo}}, \bibinfo {author} {\bibfnamefont {A.}~\bibnamefont {Ferrón}},\
  and\ \bibinfo {author} {\bibfnamefont {J.}~\bibnamefont
  {Fernández-Rossier}},\ }\bibfield  {title} {\bibinfo {title} {Remote spin
  control in haldane spin chains},\ }\href
  {https://doi.org/10.21468/SciPostPhysCore.9.2.024} {\bibfield  {journal}
  {\bibinfo  {journal} {SciPost Phys. Core}\ }\textbf {\bibinfo {volume} {9}},\
  \bibinfo {pages} {024} (\bibinfo {year} {2026})}\BibitemShut {NoStop}%
\bibitem [{\citenamefont {Lado}\ \emph {et~al.}(2017)\citenamefont {Lado},
  \citenamefont {Ferr{\'o}n},\ and\ \citenamefont
  {Fern{\'a}ndez-Rossier}}]{lado2017}%
  \BibitemOpen
  \bibfield  {author} {\bibinfo {author} {\bibfnamefont {J.~L.}\ \bibnamefont
  {Lado}}, \bibinfo {author} {\bibfnamefont {A.}~\bibnamefont {Ferr{\'o}n}},\
  and\ \bibinfo {author} {\bibfnamefont {J.}~\bibnamefont
  {Fern{\'a}ndez-Rossier}},\ }\bibfield  {title} {\bibinfo {title} {Exchange
  mechanism for electron paramagnetic resonance of individual adatoms},\
  }\href@noop {} {\bibfield  {journal} {\bibinfo  {journal} {Physical Review
  B}\ }\textbf {\bibinfo {volume} {96}},\ \bibinfo {pages} {205420} (\bibinfo
  {year} {2017})}\BibitemShut {NoStop}%
\bibitem [{\citenamefont {Ferr{\'o}n}\ \emph {et~al.}(2019)\citenamefont
  {Ferr{\'o}n}, \citenamefont {Rodr{\'\i}guez}, \citenamefont {G{\'o}mez},
  \citenamefont {Lado},\ and\ \citenamefont
  {Fern{\'a}ndez-Rossier}}]{ferron2019}%
  \BibitemOpen
  \bibfield  {author} {\bibinfo {author} {\bibfnamefont {A.}~\bibnamefont
  {Ferr{\'o}n}}, \bibinfo {author} {\bibfnamefont {S.~A.}\ \bibnamefont
  {Rodr{\'\i}guez}}, \bibinfo {author} {\bibfnamefont {S.~S.}\ \bibnamefont
  {G{\'o}mez}}, \bibinfo {author} {\bibfnamefont {J.~L.}\ \bibnamefont
  {Lado}},\ and\ \bibinfo {author} {\bibfnamefont {J.}~\bibnamefont
  {Fern{\'a}ndez-Rossier}},\ }\bibfield  {title} {\bibinfo {title} {Single spin
  resonance driven by electric modulation of the g-factor anisotropy},\
  }\href@noop {} {\bibfield  {journal} {\bibinfo  {journal} {Physical Review
  Research}\ }\textbf {\bibinfo {volume} {1}},\ \bibinfo {pages} {033185}
  (\bibinfo {year} {2019})}\BibitemShut {NoStop}%
\bibitem [{\citenamefont {Rodr{\'\i}guez}\ \emph {et~al.}(2023)\citenamefont
  {Rodr{\'\i}guez}, \citenamefont {G{\'o}mez}, \citenamefont
  {Fern{\'a}ndez-Rossier},\ and\ \citenamefont
  {Ferr{\'o}n}}]{rodriguez2023optimizing}%
  \BibitemOpen
  \bibfield  {author} {\bibinfo {author} {\bibfnamefont {S.~A.}\ \bibnamefont
  {Rodr{\'\i}guez}}, \bibinfo {author} {\bibfnamefont {S.~S.}\ \bibnamefont
  {G{\'o}mez}}, \bibinfo {author} {\bibfnamefont {J.}~\bibnamefont
  {Fern{\'a}ndez-Rossier}},\ and\ \bibinfo {author} {\bibfnamefont
  {A.}~\bibnamefont {Ferr{\'o}n}},\ }\bibfield  {title} {\bibinfo {title}
  {Optimizing tip-surface interactions in esr-stm experiments},\ }\href@noop {}
  {\bibfield  {journal} {\bibinfo  {journal} {Physical Review B}\ }\textbf
  {\bibinfo {volume} {107}},\ \bibinfo {pages} {155406} (\bibinfo {year}
  {2023})}\BibitemShut {NoStop}%
\bibitem [{\citenamefont {Phark}\ \emph
  {et~al.}(2023{\natexlab{b}})\citenamefont {Phark}, \citenamefont {Bui},
  \citenamefont {Ferr{\'o}n}, \citenamefont {Fern{\'a}ndez-Rossier},
  \citenamefont {Reina-G{\'a}lvez}, \citenamefont {Wolf}, \citenamefont {Wang},
  \citenamefont {Yang}, \citenamefont {Heinrich},\ and\ \citenamefont
  {Lutz}}]{phark2023electric}%
  \BibitemOpen
  \bibfield  {author} {\bibinfo {author} {\bibfnamefont {S.-h.}\ \bibnamefont
  {Phark}}, \bibinfo {author} {\bibfnamefont {H.~T.}\ \bibnamefont {Bui}},
  \bibinfo {author} {\bibfnamefont {A.}~\bibnamefont {Ferr{\'o}n}}, \bibinfo
  {author} {\bibfnamefont {J.}~\bibnamefont {Fern{\'a}ndez-Rossier}}, \bibinfo
  {author} {\bibfnamefont {J.}~\bibnamefont {Reina-G{\'a}lvez}}, \bibinfo
  {author} {\bibfnamefont {C.}~\bibnamefont {Wolf}}, \bibinfo {author}
  {\bibfnamefont {Y.}~\bibnamefont {Wang}}, \bibinfo {author} {\bibfnamefont
  {K.}~\bibnamefont {Yang}}, \bibinfo {author} {\bibfnamefont {A.~J.}\
  \bibnamefont {Heinrich}},\ and\ \bibinfo {author} {\bibfnamefont {C.~P.}\
  \bibnamefont {Lutz}},\ }\bibfield  {title} {\bibinfo {title}
  {Electric-field-driven spin resonance by on-surface exchange coupling to a
  single-atom magnet},\ }\href@noop {} {\bibfield  {journal} {\bibinfo
  {journal} {Advanced Science}\ }\textbf {\bibinfo {volume} {10}},\ \bibinfo
  {pages} {2302033} (\bibinfo {year} {2023}{\natexlab{b}})}\BibitemShut
  {NoStop}%
\bibitem [{\citenamefont {Phark}\ \emph
  {et~al.}(2023{\natexlab{c}})\citenamefont {Phark}, \citenamefont {Chen},
  \citenamefont {Wolf}, \citenamefont {Bui}, \citenamefont {Wang},
  \citenamefont {Haze}, \citenamefont {Kim}, \citenamefont {Lutz},
  \citenamefont {Heinrich},\ and\ \citenamefont {Bae}}]{phark2023double}%
  \BibitemOpen
  \bibfield  {author} {\bibinfo {author} {\bibfnamefont {S.-H.}\ \bibnamefont
  {Phark}}, \bibinfo {author} {\bibfnamefont {Y.}~\bibnamefont {Chen}},
  \bibinfo {author} {\bibfnamefont {C.}~\bibnamefont {Wolf}}, \bibinfo {author}
  {\bibfnamefont {H.}~\bibnamefont {Bui}}, \bibinfo {author} {\bibfnamefont
  {Y.}~\bibnamefont {Wang}}, \bibinfo {author} {\bibfnamefont {M.}~\bibnamefont
  {Haze}}, \bibinfo {author} {\bibfnamefont {J.}~\bibnamefont {Kim}}, \bibinfo
  {author} {\bibfnamefont {C.}~\bibnamefont {Lutz}}, \bibinfo {author}
  {\bibfnamefont {A.}~\bibnamefont {Heinrich}},\ and\ \bibinfo {author}
  {\bibfnamefont {Y.}~\bibnamefont {Bae}},\ }\bibfield  {title} {\bibinfo
  {title} {Double electron spin resonance of engineered atomic structures on a
  surface},\ }\href@noop {} {\bibfield  {journal} {\bibinfo  {journal} {ACS
  Nano}\ }\textbf {\bibinfo {volume} {17}},\ \bibinfo {pages} {14144} (\bibinfo
  {year} {2023}{\natexlab{c}})}\BibitemShut {NoStop}%
\bibitem [{\citenamefont {Reina-G{\'a}lvez}\ \emph {et~al.}(2025)\citenamefont
  {Reina-G{\'a}lvez}, \citenamefont {Le}, \citenamefont {Bui}, \citenamefont
  {Phark}, \citenamefont {Lorente},\ and\ \citenamefont
  {Wolf}}]{reina2025efficient}%
  \BibitemOpen
  \bibfield  {author} {\bibinfo {author} {\bibfnamefont {J.}~\bibnamefont
  {Reina-G{\'a}lvez}}, \bibinfo {author} {\bibfnamefont {H.-A.}\ \bibnamefont
  {Le}}, \bibinfo {author} {\bibfnamefont {H.~T.}\ \bibnamefont {Bui}},
  \bibinfo {author} {\bibfnamefont {S.-h.}\ \bibnamefont {Phark}}, \bibinfo
  {author} {\bibfnamefont {N.}~\bibnamefont {Lorente}},\ and\ \bibinfo {author}
  {\bibfnamefont {C.}~\bibnamefont {Wolf}},\ }\bibfield  {title} {\bibinfo
  {title} {Efficient driving of a spin qubit using single-atom magnets},\
  }\href@noop {} {\bibfield  {journal} {\bibinfo  {journal} {Physical Review
  Research}\ }\textbf {\bibinfo {volume} {7}},\ \bibinfo {pages} {L042055}
  (\bibinfo {year} {2025})}\BibitemShut {NoStop}%
\bibitem [{\citenamefont {Ferr{\'o}n}\ \emph {et~al.}(2010)\citenamefont
  {Ferr{\'o}n}, \citenamefont {Dom{\'\i}nguez},\ and\ \citenamefont
  {S{\'a}nchez}}]{ferron2010}%
  \BibitemOpen
  \bibfield  {author} {\bibinfo {author} {\bibfnamefont {A.}~\bibnamefont
  {Ferr{\'o}n}}, \bibinfo {author} {\bibfnamefont {D.}~\bibnamefont
  {Dom{\'\i}nguez}},\ and\ \bibinfo {author} {\bibfnamefont {M.~J.}\
  \bibnamefont {S{\'a}nchez}},\ }\bibfield  {title} {\bibinfo {title}
  {Large-amplitude harmonic driving of highly coherent flux qubits},\
  }\href@noop {} {\bibfield  {journal} {\bibinfo  {journal} {Physical Review
  B}\ }\textbf {\bibinfo {volume} {82}},\ \bibinfo {pages} {134522} (\bibinfo
  {year} {2010})}\BibitemShut {NoStop}%
\bibitem [{\citenamefont {Wang}\ \emph
  {et~al.}(2023{\natexlab{b}})\citenamefont {Wang}, \citenamefont {Haze},
  \citenamefont {Bui}, \citenamefont {Soe}, \citenamefont {Aubin},
  \citenamefont {Ardavan}, \citenamefont {Heinrich},\ and\ \citenamefont
  {Phark}}]{wang2023universal}%
  \BibitemOpen
  \bibfield  {author} {\bibinfo {author} {\bibfnamefont {Y.}~\bibnamefont
  {Wang}}, \bibinfo {author} {\bibfnamefont {M.}~\bibnamefont {Haze}}, \bibinfo
  {author} {\bibfnamefont {H.~T.}\ \bibnamefont {Bui}}, \bibinfo {author}
  {\bibfnamefont {W.-h.}\ \bibnamefont {Soe}}, \bibinfo {author} {\bibfnamefont
  {H.}~\bibnamefont {Aubin}}, \bibinfo {author} {\bibfnamefont
  {A.}~\bibnamefont {Ardavan}}, \bibinfo {author} {\bibfnamefont {A.~J.}\
  \bibnamefont {Heinrich}},\ and\ \bibinfo {author} {\bibfnamefont {S.-h.}\
  \bibnamefont {Phark}},\ }\bibfield  {title} {\bibinfo {title} {Universal
  quantum control of an atomic spin qubit on a surface},\ }\href@noop {}
  {\bibfield  {journal} {\bibinfo  {journal} {npj Quantum Information}\
  }\textbf {\bibinfo {volume} {9}},\ \bibinfo {pages} {48} (\bibinfo {year}
  {2023}{\natexlab{b}})}\BibitemShut {NoStop}%
\end{thebibliography}%

\end{document}